\documentclass[3p]{elsarticle}

\journal{Journal of Computational Physics}
\usepackage{lineno,hyperref}
\usepackage{amsmath}
\usepackage{array}
\usepackage{amssymb}
\usepackage{bm}
\usepackage{enumerate}
\usepackage{subcaption}
\usepackage{paralist}
\usepackage{multirow}
\usepackage[monochrome]{xcolor}
\usepackage{float}
\usepackage[section]{placeins}

\hypersetup{
	colorlinks,
	linkcolor={blue!50!black},
	citecolor={blue!50!black},
	urlcolor={blue!80!black},
	anchorcolor = {blue!80!black},
	filecolor = {blue!80!black},
	menucolor = {blue!80!black},
	runcolor = {blue!80!black}
}

\begin{document}

\begin{frontmatter}

\title{A universal diffuse interface modeling framework for surfactants}

\author[KTH]{Shahab Mirjalili}\corref{mycorrespondingauthor}
\ead{msey@kth.se}
\cortext[mycorrespondingauthor]{Corresponding author}
\address[KTH]{FLOW, Department of Engineering Mechanics, KTH Royal Institute of Technology, SE-10044 Stockholm, Sweden}
\author[ENSEIRB-Matmeca]{Mathieu Bignolles}
\ead{mbignolles@enseirb-matmeca.fr}
\address[ENSEIRB-Matmeca]{Bordeaux Polytechnic Institute, 33400 Talence, France}

\begin{abstract}
We propose a universal diffuse-interface modeling framework for surfactant transport in two-phase flows, applicable to all solubility scenarios and to any conservative phase field method. The foundation of our framework is a general three-scalar non-equilibrium model governing the surfactant concentrations in each bulk phase and at the interface, which builds on our prior consistent scalar transport framework and is locally and globally conservative, leakage-free, Galilean-invariant, and reduction-consistent. Assuming thermochemical equilibrium, we derive two one-scalar models: one for a surfactant in full equilibrium between both bulk phases and the interface, and one for a surfactant confined to a single bulk phase and the interface. All models are coupled to the Navier-Stokes equations through a surface tension force that incorporates the Marangoni stress arising from non-uniform interfacial surfactant distributions. While diffuse-interface surfactant models have been developed for the Cahn-Hilliard equation and, more recently, for the conservative Allen-Cahn (CAC) equation, existing models for CAC address only the insoluble and single-phase-soluble cases, leaving the general scenario of partial solubility in both bulk phases unaddressed; furthermore, these models lack Galilean invariance and reduction consistency, and are not applicable beyond the CAC setting. The framework is validated against analytical solutions in one-dimensional transport tests, assessed for convergence in two-dimensional advection-diffusion simulations, and demonstrated in fully-coupled drop-in-shear flow simulations covering insoluble, soluble, and partially-soluble surfactant scenarios.
\end{abstract}


\end{frontmatter}

\section{Introduction}
\label{sec:introduction}

Surfactants are amphiphilic molecules that preferentially adsorb at fluid-fluid interfaces, where they reduce the local surface tension and, when distributed non-uniformly, give rise to tangentially directed Marangoni stresses that drive flow along the interface \citep{DeGennes2003}. These properties make surfactants central to a broad range of industrial and environmental processes, including emulsification and foam stabilization, enhanced oil recovery, microfluidic drop manipulation, drug delivery, and pulmonary mechanics. The complexity of surfactant-laden flows is compounded by the wide range of solubility scenarios a given system may exhibit: depending on its molecular structure, a surfactant may be practically insoluble in the bulk phases, soluble in one of them, or partially soluble in both, with the equilibrium partitioning governed by the partition coefficient \citep{DeGennes2003}. Accurately capturing these dynamics in numerical simulations requires resolving the coupled processes of bulk advection-diffusion within each phase, adsorption and desorption between the bulk and the interface, advection and diffusion along the interface, and the feedback of the non-uniform surface tension onto the momentum balance through Marangoni stresses. Computational approaches developed for this problem span a wide range of interface-capturing frameworks, including volume-of-fluid methods \citep{James2004,Seric2018}, front-tracking methods \citep{Muradoglu2014}, and level-set approaches \citep{Khatri2011}, as well as diffuse-interface methods based on the Cahn-Hilliard equation \citep{teigen2009,Teigen2011,Soligo2019} and the conservative Allen-Cahn equation \citep{jain_surf1,jain_surf2}.

Phase field methods have emerged as a popular alternative to sharp-interface approaches for simulating two-phase flows \citep{Mirjalili_ARB,Mirjalili_comparison}, and their diffuse representation of the interface is particularly well-suited to surfactant modeling: the smooth volumetric transition region naturally accommodates bulk and interfacial concentration fields defined throughout the domain, without the need for explicit surface tracking or reconstruction even during topology changes such as drop breakup and coalescence. Among phase field models, the Cahn-Hilliard (CH) equation has historically been the primary vehicle for surfactant transport because its thermodynamic free energy formulation provides a natural basis for deriving thermodynamically consistent surfactant models \citep{Soligo2019,Zhu2019}. However, the CH equation is a fourth-order PDE, and its solutions are known to suffer from unphysical coarsening of flow features, artificial shrinkage of drops and bubbles, and a lack of bounded solutions for the phase field variable \citep{Yue2007,Mirjalili_comparison}. These difficulties have motivated the development of the conservative Allen-Cahn (CAC) equation \citep{Chiu_and_Lin}, a second-order conservative PDE that avoids the coarsening and shrinkage issues of CH, admits provably bounded solutions \citep{Mirjalili_boundedness}, and requires lower spatio-temporal resolution than CH for equivalent accuracy \citep{jain2022accurate,Mirjalili_boundedness}. A further practical advantage is that the CAC equation with $\epsilon = \Delta x$ ensures the interface is always resolved on any uniform mesh by construction, eliminating the need for the block-structured adaptive mesh refinement that CH-based solvers typically require to track the interface efficiently \citep{Teigen2011}. The CAC equation, however, lacks a thermodynamic free energy, which means that the derivation strategy used to extend CH to surfactant transport does not carry over directly, and constructing consistent, general surfactant models for CAC has remained an open problem.

An alternative approach that does not rely on a thermodynamic free energy was established by Teigen et al. \citep{teigen2009,Teigen2011}, who derived volumetric transport equations for bulk and interfacial concentrations by directly reformulating the sharp-interface surfactant equations using a smoothed Dirac delta function to represent the interface. This geometric approach does not depend on the specific form of the phase field equations and is therefore not limited to the CH setting. More recently, Jain derived surfactant transport models for the CAC equation, inspired by the geometric approach of Teigen et al.\ but arriving at a distinct formulation that is equivalent to theirs only when the phase field is at equilibrium, first for the insoluble case \citep{jain_surf1} and subsequently for a surfactant soluble in one bulk phase \citep{jain_surf2}; the general scenario in which the surfactant is partially soluble in both phases remains unaddressed. Furthermore, these models lack Galilean invariance and reduction consistency, and are specific to the CAC equation, making them inapplicable to other conservative phase field methods.

In this work, we extend the consistent scalar transport framework of \cite{mirjalili2022computational,mirjalili2022consistent} to surfactant transport. The foundation is a general three-scalar non-equilibrium model governing the surfactant concentrations in each bulk phase and at the interface, applicable to all solubility scenarios, including insoluble, soluble in one phase, and partially soluble in both phases, and to any conservative phase field method. The model is locally and globally conservative, leakage-free, positivity-preserving, and symmetric with respect to the two phases, and is consistent in the sense of \cite{mirjalili2022consistent}: it is Galilean-invariant through the inclusion of the phase field correction flux $\vec{R}$ in the transport equations, and reduction-consistent, recovering single-phase scalar transport for an artificial interface and the non-surfactant interphase transfer model of \cite{mirjalili2022consistent} in the appropriate limit. Assuming thermochemical equilibrium, we derive two one-scalar models, each governing a single concentration field: one for a surfactant in full equilibrium between both bulk phases and the interface, and one for a surfactant confined to a single bulk phase and the interface. The surfactant transport models are coupled to the Navier-Stokes equations through a surface tension force that incorporates both the normal capillary term and the Marangoni stress arising from tangential gradients of the surface tension coefficient, which depends on the local interfacial surfactant concentration through a nonlinear equation of state.

In the following, we present the surfactant transport models in Section~\ref{sec:model}, beginning with the general three-scalar non-equilibrium model and proceeding to the two one-scalar equilibrium reductions and the coupling to the Navier-Stokes equations. The numerical discretization is described in Section~\ref{sec:computational_approach}. Section~\ref{sec:results} presents numerical tests, including one-dimensional transport tests against analytical solutions, two-dimensional advection-diffusion tests with convergence studies, and fully-coupled drop-in-shear simulations covering insoluble, soluble, and partially-soluble surfactant scenarios. We summarize our findings in Section~\ref{sec:conclusions}.

\section{Models}
\label{sec:model}
Phase field methods have emerged as a popular option to capture the evolution of interfacial two-phase flows \citep{Mirjalili_ARB,garcia2025numerical}. Conservative phase field equations are the most suitable option, which can be written as
\begin{equation}
    \frac{\partial\phi}{\partial t}+\nabla\cdot(\vec{u}\phi)=\nabla\cdot\vec{R},
    \label{eqn:nphase_pf_variable_thickness}
\end{equation}
where $\phi$ is the phase field variable, which can be considered the volume fraction of one of the phases. Without loss of generality we refer to this phase as phase $1$, so the volume fraction of phase $1$ and $2$ are given by $\phi=\phi_1$ and $\phi_2=1-\phi_1$, respectively. The specific form of $\vec{R}$ determines the phase field method of choice, but the proposed models in this work for surfactant transport are broadly applicable to any conservative phase field equation, including the Cahn-Hilliard equation \citep{Cahn_Hilliard,Khanwale2020}, the conservative Allen-Cahn equation (also known as the conservative diffuse interface method) and its variants \citep{Chiu_and_Lin,Mirjalili_boundedness,jain2022accurate}. In the numerical tests of Section~\ref{sec:results}, we demonstrate the framework exclusively in the CAC setting, as this is the phase field method underlying our solver \citep{mirjalili2022computational,mirjalili2022consistent}; the extension to other conservative phase field methods is straightforward. For the CAC equation, $\vec{R} = \gamma(\epsilon\nabla\phi - \phi(1-\phi)\vec{n})$, where $\vec{n} = \nabla\phi/|\nabla\phi|$ is the interface normal, $\epsilon$ is the interface thickness parameter, and $\gamma$ is a parameter controlling the strength of the sharpening and diffusion terms \citep{Chiu_and_Lin,Mirjalili_boundedness}. 

\subsection{General three-scalar surfactant model}
\label{sec:three_scalar}
We extend the two-scalar scalar transport models introduced in \cite{mirjalili2022computational,mirjalili2022consistent} to surfactant transport. Therefore, similar to those models, the concentration of the surfactant in the bulks of phase $1$ and $2$ per total volume, are denoted by $c_1$ and $c_2$, respectively. To account for surfactants, we introduce an additional variable, the interfacial surfactant concentration per total volume, denoted by $s$. The interface is converted to a volumetric representation by a smoothened Dirac function that indicates the interface, $\delta$. This is similar to previous work in the literature \citep{teigen2009,Teigen2011,jain_surf1,jain_surf2}. Throughout the one-dimensional and two-dimensional tests in this work we use $\delta=\phi_1\phi_2/\epsilon$; the flow-coupled tests in Section~\ref{sec:flow_coupled} use $\delta=6\phi_1^2\phi_2^2/\epsilon$ to match the convention of \cite{Teigen2011}. Corresponding to $c_p$ and $s$ which are concentrations per total volume in phase $p=1, 2$, and the interface, respectively, the phasic concentrations $\tilde{c}_p=c_p/\phi_p$, defined as the concentration of the surfactant in phase $p$ per unit volume of phase $p$, and the specific interfacial concentration of the surfactant, $\tilde{s}=s/\delta$ are important field variables that are used in flux calculation.

We propose the following general three-scalar transport model for $c_1$, $c_2$, and $s$, which is applicable to transport of surfactants in any two-phase setting (e.g., insoluble, soluble in both phases, soluble in one phase):
\begin{equation}
    \frac{\partial c_1}{\partial t}+\nabla\cdot(\vec{u}c_1)=\nabla\cdot(D_1 \phi_1\nabla(\frac{c_1}{\phi_1}))+\nabla\cdot(\vec{R}\frac{c_1}{\phi_1})+J_{21}+J_{s1}\delta,
 \end{equation}
 \begin{equation}
    \frac{\partial c_2}{\partial t}+\nabla\cdot(\vec{u}c_2)=\nabla\cdot(D_2 \phi_2\nabla(\frac{c_2}{\phi_2}))+\nabla\cdot(-\vec{R}\frac{c_2}{\phi_2})+J_{12}+J_{s2}\delta,
 \end{equation}
 \begin{equation}
     \frac{\partial s}{\partial t}+\nabla\cdot(\vec{u}s)=\nabla\cdot(D_s \delta\nabla(\frac{s}{\delta}))+J_{1s}\delta+J_{2s}\delta
 \end{equation}
 
As derived in \cite{mirjalili2022computational}, the transfer term from phase $2$ to phase $1$ is given by 
\begin{equation}
    J_{12}=A D_{m}(K_{eq}c_2\phi_1-c_1\phi_2)-D_{m}\nabla\phi_1\cdot\nabla(c_1+K_{eq}c_2),
\end{equation}
where the mixed diffusivity is
\begin{equation}
    D_{m}=\frac{D_1D_2}{K_{eq}D_1\phi_2+D_2\phi_1}.
\end{equation}
At chemical equilibrium, a jump in concentration is enforced across the interface, specified by the partition coefficient, $K_{eq}$, 
\begin{equation}
    K_{eq}=(\frac{\tilde{c}_1}{\tilde{c}_2})_\text{eq}.
    \label{eqn:K_eq}
\end{equation}
The transfer term from phase $1$ to phase $2$ is $J_{12}=-J_{21}$, which can be confirmed by inspection if one switches the numbering of the two phases.

The transfer term from the bulk of phase $p=1,2$ to the interface is $J_{ps}\delta$, where
\begin{equation}
    J_{ps}=r_{a,p}\frac{c_p}{\phi_p}(\tilde{s}_{\infty}-\frac{s}{\delta})-r_{d,p}\frac{s}{\delta}.
    \label{eqn:surf_transfer}
\end{equation}
The adsorption and desorption coefficients between the bulks of phase $p$ and the interface are $r_{a,p}$ and $r_{d,p}$ respectively. Moreover, the saturation interfacial concentration is denoted by $\tilde{s}_\infty$. Naturally, the transfer term from the interface to the bulk of phase $p$ is $J_{sp}\delta=-J_{ps}\delta$. Surfactant transport equilibrium with zero transfer and diffusion fluxes can only be achieved if
\begin{equation}
    \frac{r_{d,p}}{r_{a,p}}=K_{pq}\frac{r_{d,q}}{r_{a,q}}.
\end{equation}

The proposed three-scalar model satisfies several important properties. It is locally and globally conservative: summing the three transport equations, the inter-phase transfer terms cancel pairwise, since $J_{12} = -J_{21}$, $J_{1s} = -J_{s1}$, and $J_{2s} = -J_{s2}$, so that the total surfactant amount $c_1 + c_2 + s$ is conserved in a closed domain. The model is also universal with respect to the solubility scenario. The insoluble case is recovered by setting $D_p = r_{a,p} = r_{d,p} = 0$ for both phases, leaving only the interfacial equation for $s$ active. A surfactant soluble only in phase $p$ follows from $D_q = r_{a,q} = r_{d,q} = 0$ for the insoluble phase $q \neq p$. In the non-surfactant limit ($r_{a,p} = r_{d,p} = 0$, $s = 0$), the equations for $c_1$ and $c_2$ reduce exactly to the interphase transfer model of \cite{mirjalili2022consistent}, establishing reduction consistency with the scalar transport framework this work extends. The case of an artificial interface in a single-phase flow, examined in Section~\ref{sec:results}, recovers single-phase scalar transport exactly. By construction, the model inherits the leakage-free property of \cite{mirjalili2022computational,mirjalili2022consistent}: in the absence of physical adsorption and desorption, the phase-weighted diffusion terms prevent unphysical transfer of surfactant across the interface even for large inter-phase concentration contrasts. Positivity of $c_1$, $c_2$, and $s$ is likewise expected to be preserved, following the same structural reasoning as in \cite{mirjalili2022computational,mirjalili2022consistent}.

The model is also agnostic to the specific conservative phase field method employed: $\phi_1$, $\phi_2$, $\vec{R}$, and $\delta$ appear in the surfactant transport equations, while their evolution is determined by whichever conservative phase field equation is adopted (Eq.~\eqref{eqn:nphase_pf_variable_thickness}). This makes the framework applicable without modification to the Cahn-Hilliard equation \citep{Cahn_Hilliard,Khanwale2020}, the conservative Allen-Cahn equation \citep{Chiu_and_Lin,Mirjalili_boundedness,jain2022accurate}, and any other conservative formulation. The consistent inclusion of the phase field correction flux $\vec{R}$ in the scalar transport equations, following \cite{mirjalili2022consistent}, also ensures Galilean invariance; the solver described in Section~\ref{sec:computational_approach} preserves these properties at the discrete level.

The three-scalar model introduced in this section can be used for simulating the transport of surfactants in any setting (e.g., liquid-liquid, liquid-gas), including when the interfaces have not achieved thermochemical equilibrium. As demonstrated in \cite{mirjalili2022computational} for the case of passive scalar transport in two-phase systems, a one-scalar transport model can be derived by assuming thermochemical equilibrium. In the following, we will make such equilibrium assumptions to derive one-scalar models for surfactant transport.

\subsection{One-scalar model for surfactant soluble in both phases (full equilibrium)}
 We derive a one-scalar model for surfactant transport by assuming chemical equilibrium, which means that not only is Eq. \eqref{eqn:K_eq} satisfied, resulting in $J_{pq}=0$, but $J_{ps}=0$ for any value of $p$ and $q$. The one-scalar equation governing surfactant concentration, $c$, is
\begin{equation}
    \frac{\partial c}{\partial t}+\nabla\cdot(\vec{u}c)=\nabla\cdot(D_\text{eff}\nabla(\frac{c_{b}}{K_\text{eff}}))+\nabla\cdot(\vec{R}(K_{eq}-1)\frac{c_b}{K_\text{eff}})+\nabla\cdot(D_s \delta\nabla(\frac{c-c_b}{\delta})),
    \label{eqn:one_scalar}
\end{equation}
where 
\begin{equation}
    D_\text{eff}=D_1K_{eq}\phi_1+D_2\phi_2,
\end{equation}
and
\begin{equation}
    K_\text{eff}=K_\text{eq}\phi_1+\phi_2,
\end{equation}
are the effective mixture diffusivity and Henry coefficient, respectively. In Eq.~\eqref{eqn:one_scalar}, $c_b$ is the concentration in the bulk, given by
\begin{equation}
    c_b=\frac{-f+\sqrt{f^2+4r_{a,1}r_{d,1}K_\text{eq}K_\text{eff}c}}{2r_{a,1}K_\text{eq}},
\end{equation}
where 
\begin{equation}
 f=r_{d,1}K_\text{eff}+r_{a,1}K_\text{eq}\tilde{s}_\infty\delta-r_{a,1}K_\text{eq}c.
\end{equation}

\subsection{One-scalar model for surfactant soluble in one phase (confined scalar)}
A common scenario is when the surfactant is insoluble or practically insoluble in one of the phases. Then, it is only reasonable to assume equilibrium between the surfactant concentration in the bulk of the soluble phase and the interface. Without loss of generality, let us assume that the soluble phase is phase $1$. Then, the following one-scalar model can be derived to solve for the surfactant concentration confined to phase $1$ and the interface, $c$, 
\begin{equation}
    \frac{\partial c}{\partial t}+\nabla\cdot(\vec{u}c)=\nabla\cdot(D_1\phi_1\nabla(\frac{c_{1}}{\phi_1}))+\nabla\cdot(\vec{R}\frac{c_1}{\phi_1})+\nabla\cdot(D_s \delta\nabla(\frac{c-c_1}{\delta})),
\end{equation}
where $c_1$ is the concentration in the bulk of phase $1$ per total volume, given by
\begin{equation}
    c_1=\frac{-f+\sqrt{f^2+4r_{a,1}r_{d,1}\phi_1c}}{2r_{a,1}},
\end{equation}
where 
\begin{equation}
    f=r_{d,1}\phi_1+r_{a,1}\tilde{s}_\infty\delta-r_{a,1}c.
\end{equation}

\subsection{Coupling to Navier-Stokes}
These equations are coupled to the Navier-Stokes equation,
\begin{equation}
    \frac { \partial (\rho \vec{ u } ) }{ \partial t } +\nabla \cdot \left[(\rho \vec{ u }-\vec{S} )\otimes \vec{ u } \right ]=-\nabla P+\nabla \cdot \left[\mu (\nabla \vec{ u } +\nabla ^{ T }\vec{ u } ) \right ]+\vec{ F }_\text{ ST }.
    \label{eqn:momentum}
\end{equation}
In Eq.~\eqref{eqn:momentum}, $P$ is the pressure field, $\rho$ is the density field computed via
\begin{equation}
    \rho=\Sigma_{p}\rho_p\phi_p,
    \label{eqn:density}
\end{equation}
$\mu$ is the fluid viscosity, computed via
\begin{equation}
    \mu=\Sigma_{p}\mu_p\phi_p,
    \label{eqn:viscosity}
\end{equation}
and $\vec{S}$ is the mass-momentum consistency correction flux, given by 
\begin{equation}
    \vec{S}=\Sigma_{p}{\rho_p\vec{R}_p},
    \label{eqn:S_def}
\end{equation}

The surface tension force in Eq.\eqref{eqn:momentum}, $\vec{ F }_\text{ ST }$ is modeled by augmenting the CSF surface tension formulation with the Marangoni term,
\begin{equation}
    \vec{F}_\text{ST}=\sigma\kappa\nabla\phi_1+\delta\nabla_s\sigma,
    \label{eqn:ST}
\end{equation}
where $\sigma$ is the surface tension coefficient, computed via
\begin{equation}
    \sigma=\sigma_0(1+Ma\ln(1-\frac{\tilde{s}}{\tilde{s}_\infty})),
\end{equation}
where $\nabla_s=(I-\vec{n}\otimes\vec{n})\nabla$ is the surface gradient operator and $\kappa$ is the curvature computed using normal vector, 
\begin{equation}
    \kappa=-\nabla\cdot\vec{n}.
    \label{eqn:curvature}
\end{equation}
The normal vector is computed via 
\begin{equation}
    \vec{n}=\frac{\nabla\phi_1}{|\nabla\phi_1|}.
\end{equation}
The first term in Eq.~\eqref{eqn:ST} is the continuum surface force (CSF) formulation of the normal surface tension force \citep{Mirjalili2023}, while the second term, $\delta\nabla_s\sigma$, is the Marangoni stress arising from tangential gradients in surface tension along the interface. The surface tension coefficient $\sigma$ depends on the local interfacial surfactant concentration $\tilde{s} = s/\delta$ through the equation of state above: regions of elevated surfactant loading experience locally reduced $\sigma$, which drives tangential Marangoni flow toward regions of higher surface tension. When the surfactant distribution is spatially uniform, $\nabla_s\sigma = 0$ and Eq.~\eqref{eqn:ST} reduces to the standard two-phase surface tension formulation. The surfactant transport equations of Section~\ref{sec:three_scalar} and the Navier-Stokes system (Eqs.~\eqref{eqn:momentum}--\eqref{eqn:ST}) are therefore two-way coupled: the flow advects and redistributes the surfactant distribution, and the resulting non-uniform $\sigma$ feeds back into the momentum balance through the Marangoni term.

\subsection{Model selection}
\label{sec:model_selection}
The choice of model depends on both the physical solubility scenario and the degree to which the system is near thermochemical equilibrium. The solubility scenario is a property of the surfactant chemistry and the phase pair, characterized by the hydrophilic-lipophilic balance (HLB) and the partition coefficient $K_\text{eq}$ \citep{DeGennes2003}. For gas-liquid systems, the gas phase does not dissolve surfactant in practice, so the surfactant is effectively confined to the liquid phase and the interface; the confined one-scalar model or the insoluble limit ($r_{a,p} = r_{d,p} = 0$) are the appropriate choices. For liquid-liquid systems, the solubility scenario depends on $K_\text{eq}$. Surfactants with low HLB are lipophilic and partition strongly into the oil phase ($K_\text{eq} \gg 1$), making them effectively confined there; surfactants with high HLB are hydrophilic and partition into the water phase ($K_\text{eq} \ll 1$). In both cases, the confined one-scalar model is appropriate. Surfactants with intermediate HLB values, such as those used in foaming and wetting applications, exhibit significant solubility in both phases and require the partially soluble three-scalar model or its full-equilibrium reduction \citep{DeGennes2003}.

The choice between the three-scalar and one-scalar models is governed by the degree of departure from thermochemical equilibrium, characterized by the Biot number $Bi = r_{d,p}/\dot{\gamma}$ (or an equivalent ratio of the adsorption/desorption timescale to the relevant flow timescale). When $Bi \gg 1$, exchange between the bulk and the interface is fast relative to the flow, the system remains near equilibrium, and the one-scalar reduction is appropriate. When $Bi \sim O(1)$ or $Bi \ll 1$, the exchange kinetics are comparable to or slower than the flow timescale, and the three-scalar model is necessary to capture the transient non-equilibrium dynamics. The numerical tests of Section~\ref{sec:results} illustrate these distinctions in both one-dimensional transport and fully-coupled flow settings.

\section{Computational approach}
\label{sec:computational_approach}
We use a fourth-order Runge-Kutta (RK4) explicit scheme for time integration. At each time step, the phase field equation, Eq.~\eqref{eqn:nphase_pf_variable_thickness}, is integrated to update $\phi$, and the surfactant transport equations are integrated to update $c_1$, $c_2$, and $s$. These are coupled to the momentum equation, Eq.~\eqref{eqn:momentum}, through the density, viscosity, and surface tension force, which are updated from $\phi$ and $\tilde{s}$ at each stage. In space, we use a second-order central finite difference discretization on a standard staggered Cartesian grid, in which velocity components and all fluxes are stored on their respective cell faces, while pressure, density, viscosity, the phase field variable, and all concentration fields are stored at cell centers. The surfactant transport equations are discretized using the same stencils as the scalar transport framework of \cite{mirjalili2022computational}, to which the adsorption, desorption, and interfacial diffusion terms are added. The implementation is three-dimensional and parallelized using MPI.

\section{Numerical tests}
\label{sec:results}
In this section, we assess the accuracy and convergence properties of the proposed surfactant transport models through a hierarchy of test cases. We first validate the three-scalar model and its one-scalar reductions against analytical solutions for one-dimensional surfactant transport in Section~\ref{sec:uncoupled}, covering partially soluble, single-phase-soluble, and insoluble scenarios, as well as the artificial-interface limit. Convergence in two-dimensional problems is assessed in Section~\ref{sec:2D}. Finally, Section~\ref{sec:flow_coupled} presents fully-coupled drop-in-shear flow simulations covering insoluble, single-phase-soluble, and partially-soluble surfactants, demonstrating the framework's ability to capture Marangoni-driven deformation and bulk-interface exchange. Although the model and solver are formulated and implemented in three dimensions, all tests presented here are one- or two-dimensional; the lower-dimensional setting is sufficient for the purposes of model verification and validation. As noted in Section~\ref{sec:model}, all tests use the CAC equation as the underlying phase field method. All values reported are in non-dimensional form.

\subsection{One-dimensional surfactant transport tests}
\label{sec:uncoupled}
All one-dimensional tests are performed on a static domain $x \in [-1, 1]$ with $N_x = 129$ uniformly spaced cells ($\Delta x = 1/64$). The conservative Allen--Cahn equation is used with $\epsilon = \Delta x$ and $\gamma = 1$. Phase 1 occupies a drop of diameter $0.5$ centered at the origin, described by the hyperbolic tangent profile $\phi(x,0) = \frac{1}{2}(1+\tanh\frac{x+0.25}{2\epsilon}) - \frac{1}{2}(1+\tanh\frac{x-0.25}{2\epsilon})$. The velocity is zero in all cases, periodic boundary conditions are applied, and the parameter $A$ in $J_{12}$ is set to $1000$. In each test, the three-scalar model is run alongside the applicable one-scalar model(s) to assess the accuracy of the equilibrium reduction.

\subsubsection{Partially soluble surfactant}
\label{sec:1d_case1}
\begin{table}[H]
\centering
\renewcommand{\arraystretch}{1.3}
\begin{tabular}{|>{\raggedright\arraybackslash}p{4cm}
                |>{\raggedright\arraybackslash}p{3.5cm}
                |>{\raggedright\arraybackslash}p{4cm}
                |>{\raggedright\arraybackslash}p{3.5cm}|}
\hline
 & \textbf{Phase 1} & \textbf{Phase 2} & \textbf{Interface} \\ \hline

\textbf{Initial conditions ($t = 0$)}
& $\tilde{c}_1 = 2.5$
& $\tilde{c}_2 = 0.8$
& $\tilde{s} = 0$ \\ \hline

\textbf{Surfactants parameters}
& $D_1 = 1.0$, $r_{a1} = 3.0$, $r_{d1} = 1.0$
& $D_2 = 1.0$, $r_{a2} = 1.0$, $\textcolor{blue}{\boldsymbol{r_{d2}}} = \frac{1}{3}$
& $D_s = 0.2$, $\tilde{s}_{\infty} = 1.0$ \\ \hline
\end{tabular}
\caption{Simulation parameters for the test in Section~\ref{sec:1d_case1}. The partition coefficient is $K_\text{eq} = r_{d,1}r_{a,2}/(r_{a,1}r_{d,2}) = 1$.}
\label{tab1}
\end{table}

\begin{figure}[H]
    \captionsetup{labelfont=bf,font=footnotesize}
    \centering

    \begin{subfigure}[b]{0.49\textwidth}
        \includegraphics[width=\textwidth]{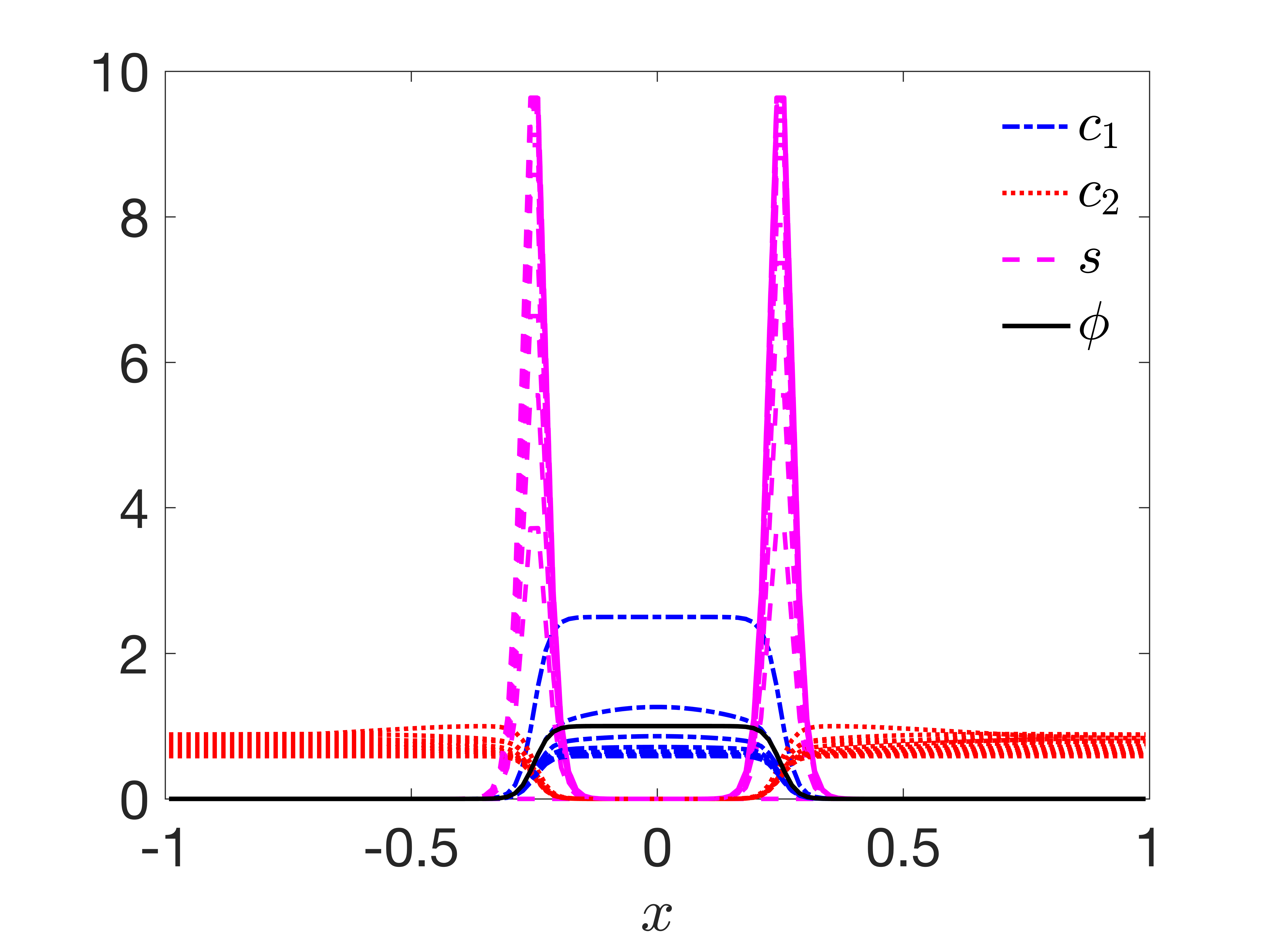}
        \caption{}
    \end{subfigure}
    \hfill
    \begin{subfigure}[b]{0.49\textwidth}
        \includegraphics[width=\textwidth]{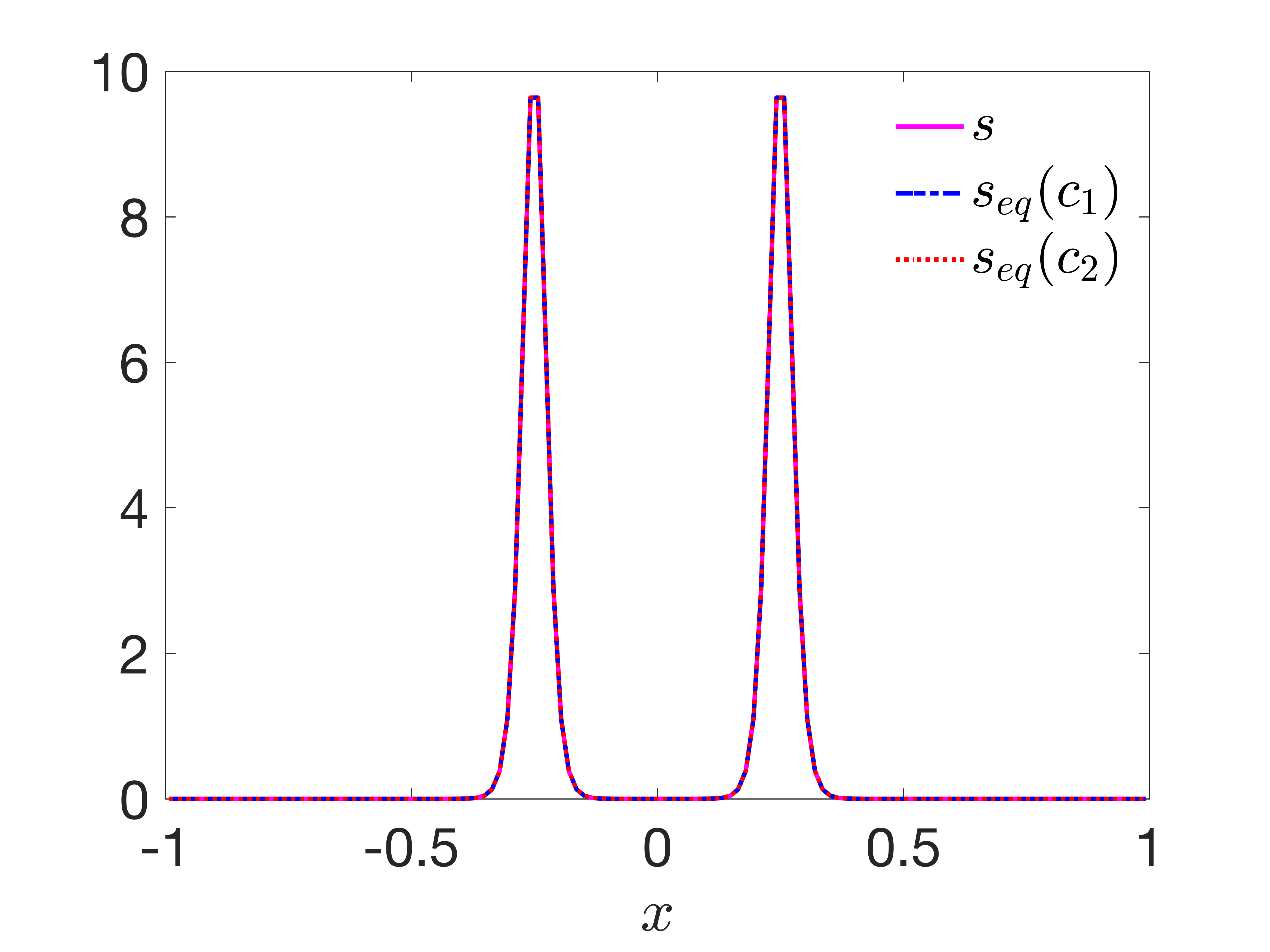}
        \caption{}
    \end{subfigure}

    \begin{subfigure}[b]{0.49\textwidth}
        \includegraphics[width=\textwidth]{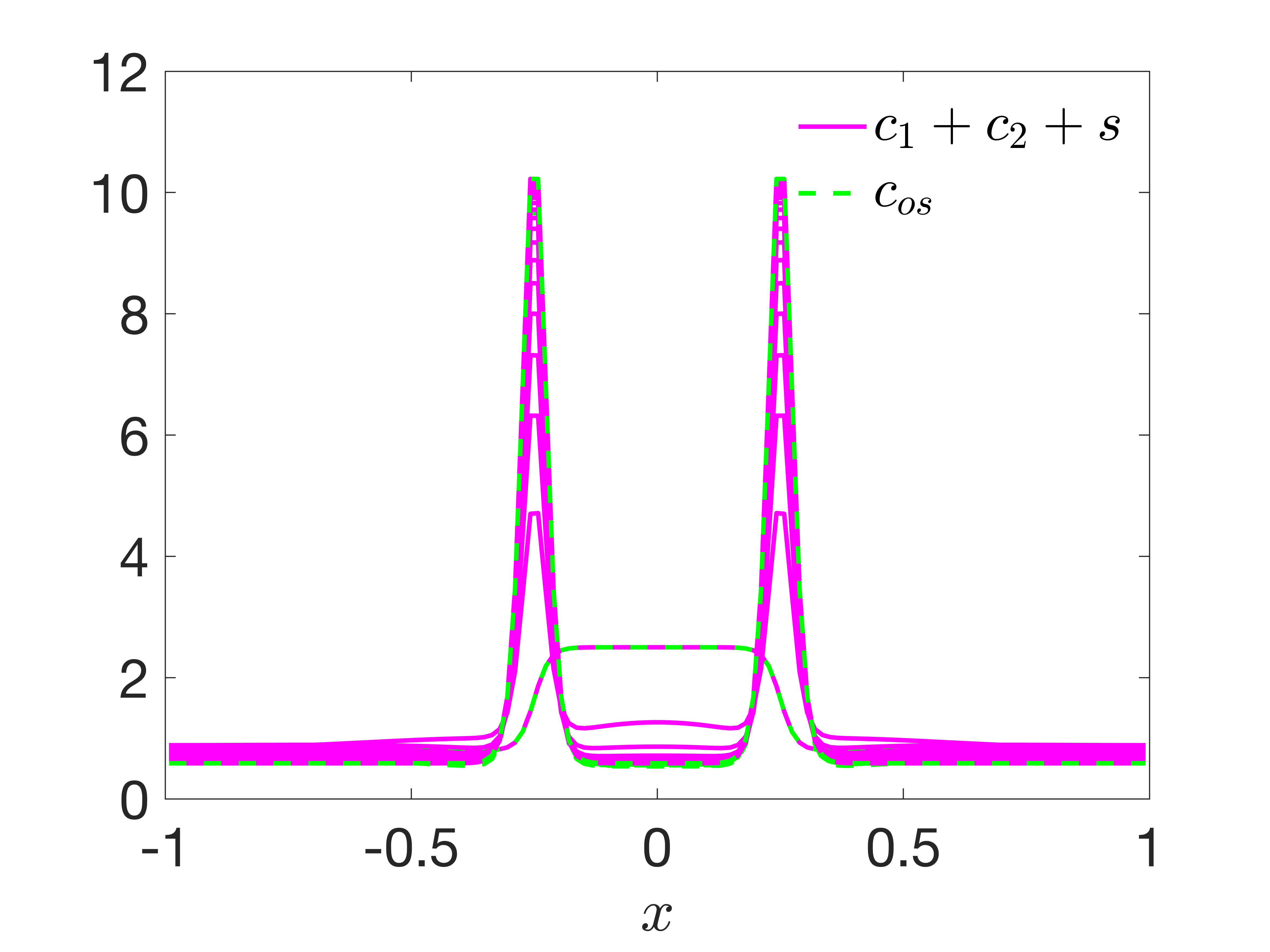}
        \caption{}
    \end{subfigure}
    \hfill
    \begin{subfigure}[b]{0.49\textwidth}
        \includegraphics[width=\textwidth]{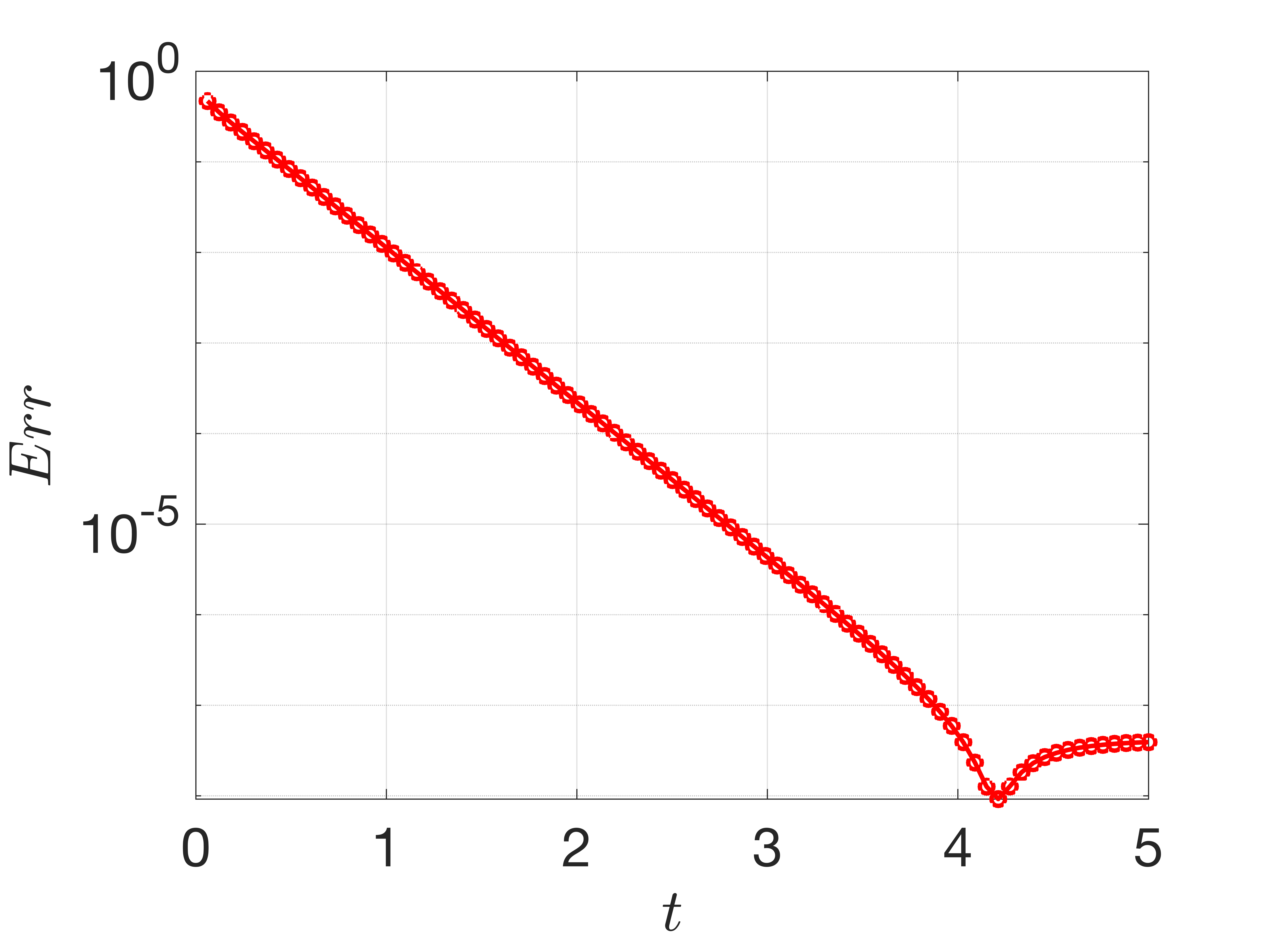}
        \caption{}
    \end{subfigure}

    \caption{One-dimensional transport for a surfactant soluble in both phases (parameters in Table~\ref{tab1}): (a) profiles of $c_1$, $c_2$, $s$, and $\phi$ at multiple time instants from the three-scalar model, (b) interfacial concentration $s$ at $t=5$ compared against the equilibrium Langmuir profiles $s_\text{eq}(\tilde{c}_1)$ and $s_\text{eq}(\tilde{c}_2)$, (c) total concentration $c_1+c_2+s$ from the three-scalar model overlaid with the full-equilibrium one-scalar model $c_\text{os}$ at multiple times, and (d) relative $L^1$ error between the two models as a function of time.}
    \label{fig:1d_case1}
\end{figure}

The simulation parameters are listed in Table~\ref{tab1}. The system is initialized far from equilibrium with $c_1(x,0) = 2.5\phi$, $c_2(x,0) = 0.8(1-\phi)$, and $s(x,0) = 0$, and integrated to $t = 5$. Figure~\ref{fig:1d_case1}(a) shows the temporal evolution of $c_1$, $c_2$, and $s$ alongside the phase field profile $\phi$: starting from the initial out-of-equilibrium state, the concentration fields relax toward equilibrium and $s$ builds up from zero as the surfactant adsorbs to the interface. At equilibrium, setting $J_{ps} = 0$ in Eq.~\eqref{eqn:surf_transfer} yields the Langmuir isotherm, $\tilde{s}_\text{eq}(\tilde{c}_p) = r_{a,p}\tilde{s}_\infty\tilde{c}_p/(r_{d,p} + r_{a,p}\tilde{c}_p)$; panel (b) confirms that the numerical $s$ at $t = 5$ agrees with both $s_\text{eq}(\tilde{c}_1)$ and $s_\text{eq}(\tilde{c}_2)$, as required by thermochemical consistency. In panel (c), the total surfactant $c_1+c_2+s$ from the three-scalar model is compared against the full-equilibrium one-scalar model $c_\text{os}$ at multiple time instants; the two converge as equilibrium is approached. The relative $L^1$ error between the two models, plotted in panel (d), decays with time, confirming that the three-scalar model converges to the equilibrium prediction of the one-scalar model as the system approaches equilibrium.

\subsubsection{Partially soluble surfactant with large diffusivity ratio}
\label{sec:1d_case2}
\begin{table}[H]
\centering
\renewcommand{\arraystretch}{1.3}
\begin{tabular}{|>{\raggedright\arraybackslash}p{4cm} 
                |>{\raggedright\arraybackslash}p{3.5cm} 
                |>{\raggedright\arraybackslash}p{4cm} 
                |>{\raggedright\arraybackslash}p{3.5cm}|}
\hline
 & \textbf{Phase 1} & \textbf{Phase 2} & \textbf{Interface} \\ \hline
\textbf{Initial conditions (at $t = 0$)}
& $\tilde{c}_1 = 2.0$
& $\tilde{c}_2 = 0$
& $\tilde{s} = 0.3$ \\ \hline

\textbf{Surfactants parameters}
& $D_1 = 1.0$, $r_{a1} = 3.0$, $r_{d1} = 1.0$
& $D_2 = 10^{-6}$, $r_{a2} = 10^{-6}$, $\textcolor{blue}{\boldsymbol{r_{d2}}} = \frac{10^{-6}}{3}$
& $D_s = 0.2$, $\tilde{s}_{\infty} = 1.0$ \\ \hline
\end{tabular}
\caption{Simulation parameters for the test in Section~\ref{sec:1d_case2}. The partition coefficient is $K_\text{eq} = 1$.}
\label{tab2}
\end{table}

\begin{figure}[H]
    \captionsetup{labelfont=bf,font=footnotesize}
    \centering

    \begin{subfigure}[b]{0.32\textwidth}
        \includegraphics[width=\textwidth]{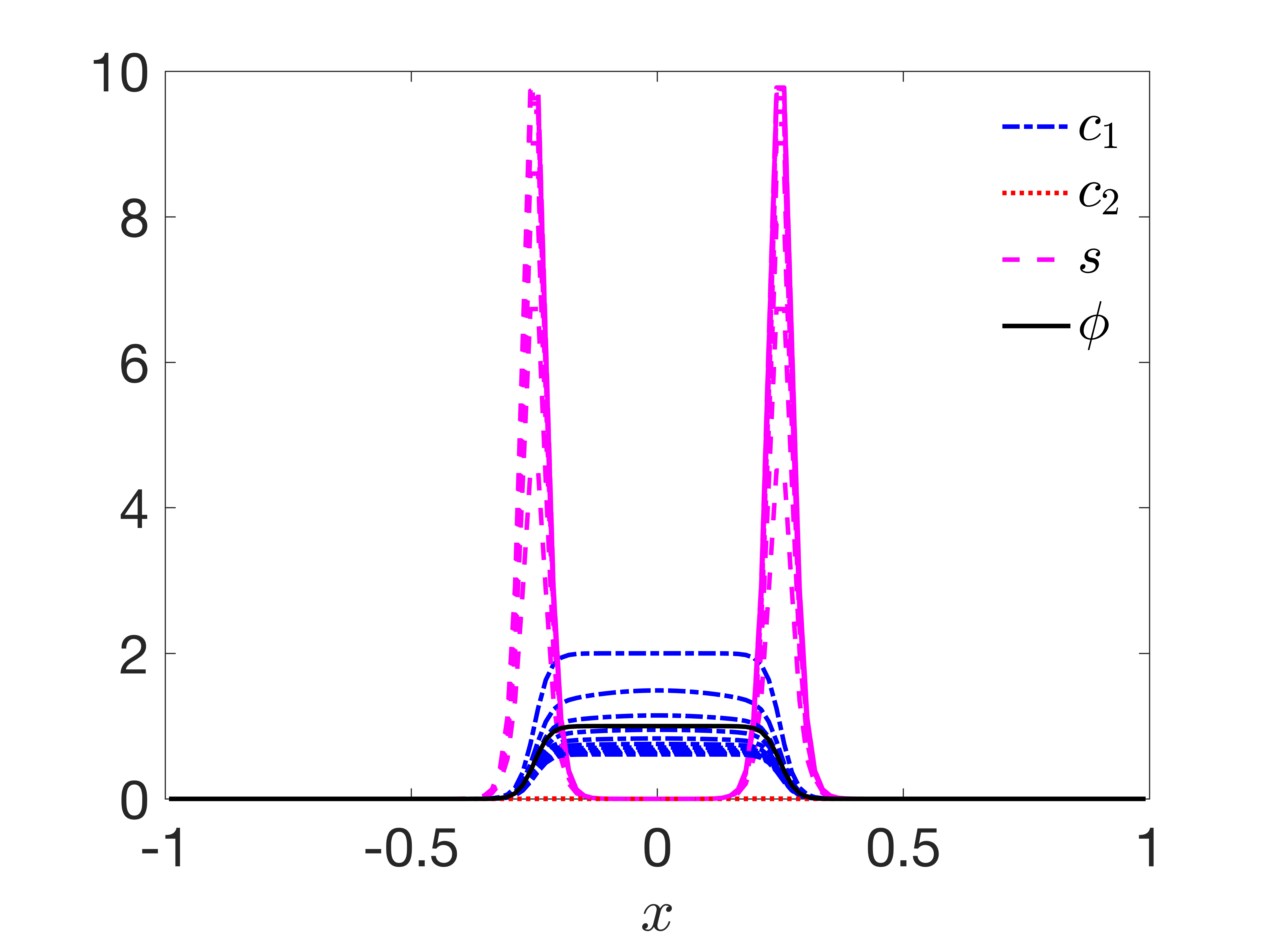}
        \caption{}
    \end{subfigure}
    \hfill
    \begin{subfigure}[b]{0.32\textwidth}
        \includegraphics[width=\textwidth]{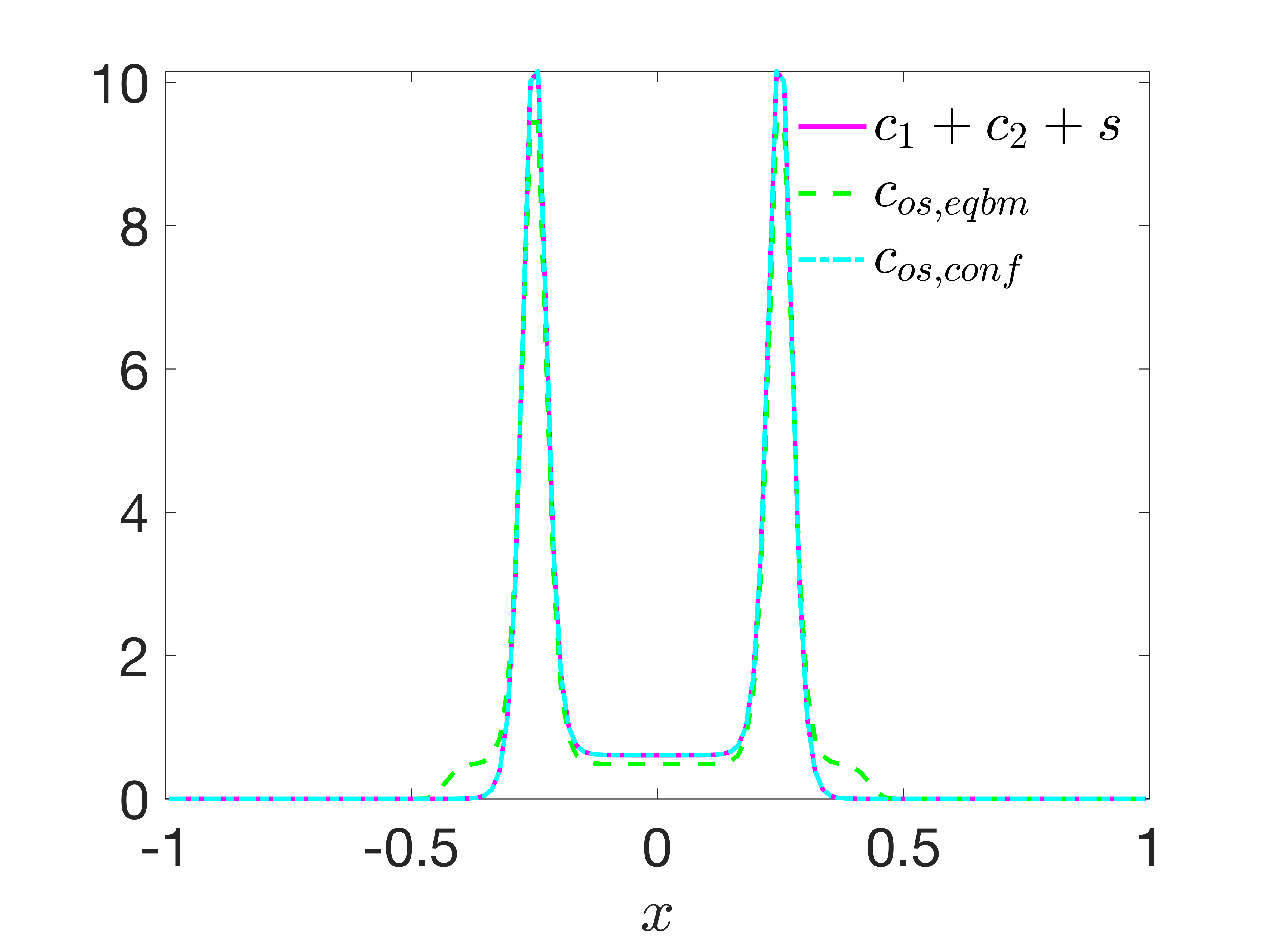}
        \caption{}
    \end{subfigure}
    \begin{subfigure}[b]{0.32\textwidth}
        \includegraphics[width=\textwidth]{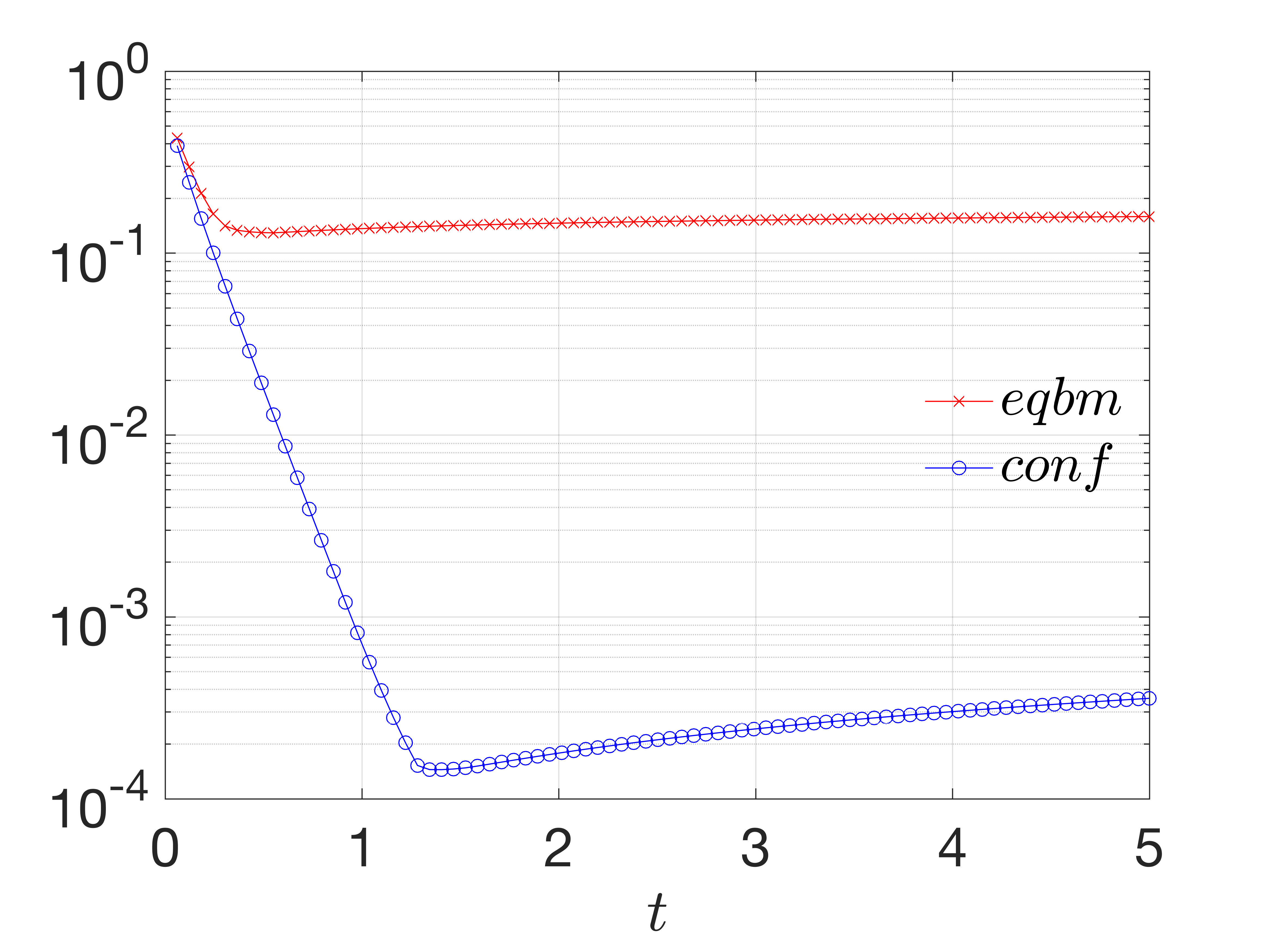}
        \caption{}
    \end{subfigure}

    \caption{One-dimensional transport for a surfactant soluble in both phases with a large diffusivity ratio $D_1/D_2 = 10^6$ (parameters in Table~\ref{tab2}): (a) profiles of $c_1$, $c_2$, $s$, and $\phi$ at multiple time instants from the three-scalar model, (b) total concentration $c_1+c_2+s$ at $t=5$ compared against the full-equilibrium one-scalar model $c_\text{os,eqbm}$ and the confined one-scalar model $c_\text{os,conf}$, and (c) relative $L^1$ errors of the two one-scalar models as functions of time.}
    \label{fig:1d_case2}
\end{figure}

The simulation parameters are listed in Table~\ref{tab2}. This test examines the case in which the surfactant is partially soluble in both phases but with a large diffusivity ratio $D_1/D_2 = 10^6$. The adsorption and desorption rates in phase 2 are scaled proportionally so that $K_\text{eq} = 1$ is preserved. The system is initialized with $c_1(x,0) = 2.0\phi$, $c_2(x,0) = 0$, and $s(x,0) = 0.3\,\delta$, and run to $t = 5$. Figure~\ref{fig:1d_case2}(a) shows the time evolution of the concentrations: surfactant in phase 1 redistributes and adsorbs to the interface, while phase 2 remains nearly depleted throughout because its diffusivity is too low for significant bulk transport. Panel (b) compares the total $c_1+c_2+s$ at $t = 5$ from the three-scalar model against both one-scalar models. The full-equilibrium model $c_\text{os,eqbm}$ substantially over-predicts the concentration in phase 2, as it enforces equilibrium between both bulk phases regardless of their transport timescales. The three-scalar result is in close agreement with the confined model $c_\text{os,conf}$, which assumes equilibrium only between phase 1 and the interface. Panel (c) confirms this quantitatively: the $L^1$ difference between the three-scalar and confined models remains small throughout the simulation, while that between the three-scalar and full-equilibrium models is large.

\subsubsection{Surfactant soluble in one phase}
\label{sec:1d_case3}
\begin{table}[h!]
\centering 
\renewcommand{\arraystretch}{1.3}
\begin{tabular}{|>{\raggedright\arraybackslash}p{4cm} 
                |>{\raggedright\arraybackslash}p{3.5cm} 
                |>{\raggedright\arraybackslash}p{4cm} 
                |>{\raggedright\arraybackslash}p{3.5cm}|}
\hline
 & \textbf{Phase 1} & \textbf{Phase 2} & \textbf{Interface} \\ \hline
\textbf{Initial conditions (at $t = 0$)}
& $\tilde{c}_1 = 2.0$
& $\tilde{c}_2 = 0$
& $\tilde{s} = 1.5$ \\ \hline

\textbf{Surfactants parameters}
& $D_1 = 0.5$, $r_{a1} = 3.0$, $r_{d1} = 2.0$
& $D_2 = 0$, $r_{a2} = 0$, $\textcolor{blue}{\boldsymbol{r_{d2}}} = 0$
& $D_s = 0.5$, $\tilde{s}_{\infty} = 1$ \\ \hline
\end{tabular}
\caption{Simulation parameters for the test in Section~\ref{sec:1d_case3}. Phase 2 is inert ($D_2 = r_{a2} = r_{d2} = 0$).}
\label{tab3}
\end{table}

\begin{figure}[H]
    \captionsetup{labelfont=bf,font=footnotesize}
    \centering

    \begin{subfigure}[b]{0.32\textwidth}
        \includegraphics[width=\textwidth]{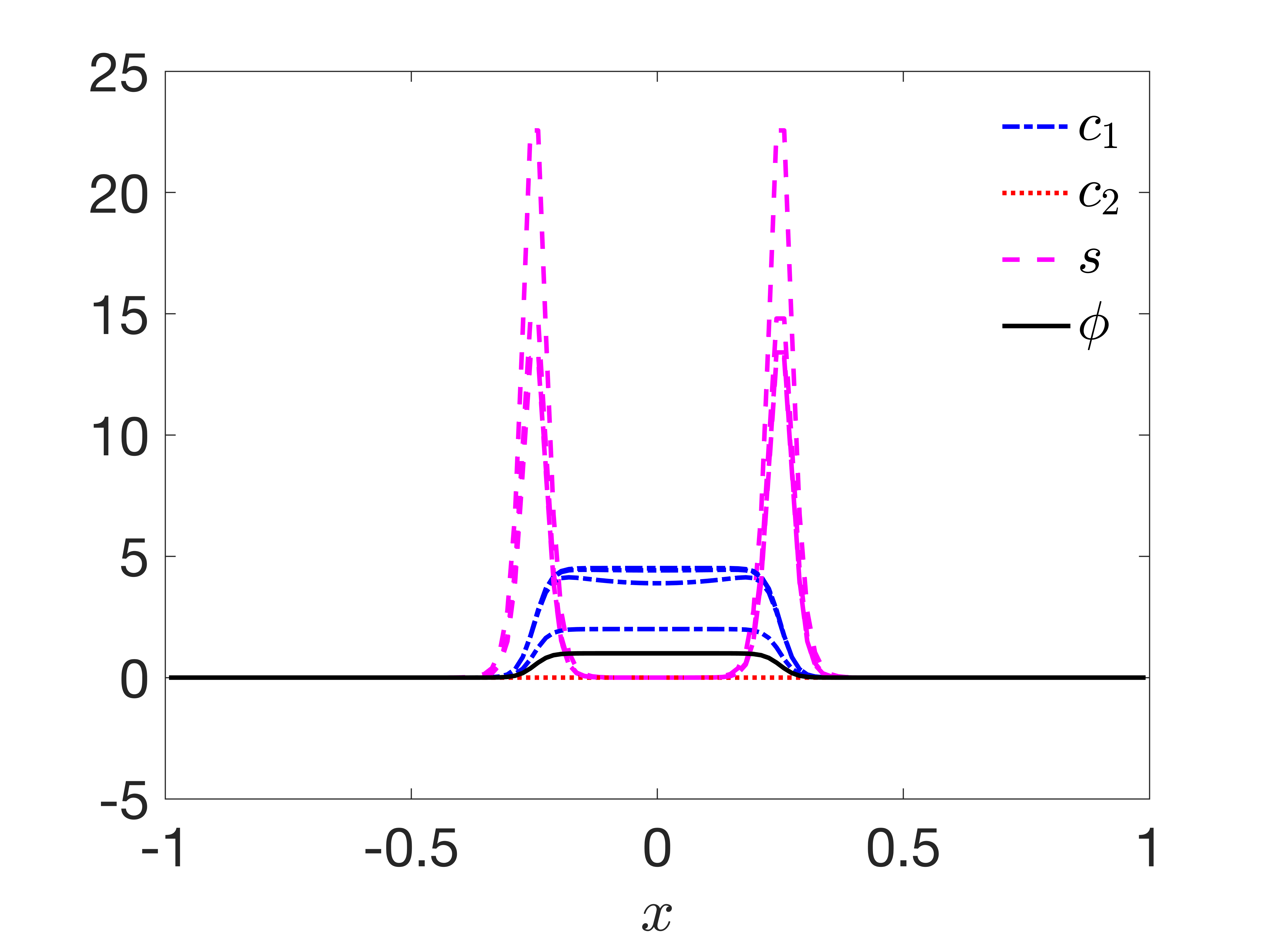}
        \caption{}
    \end{subfigure}
    \hfill
    \begin{subfigure}[b]{0.32\textwidth}
        \includegraphics[width=\textwidth]{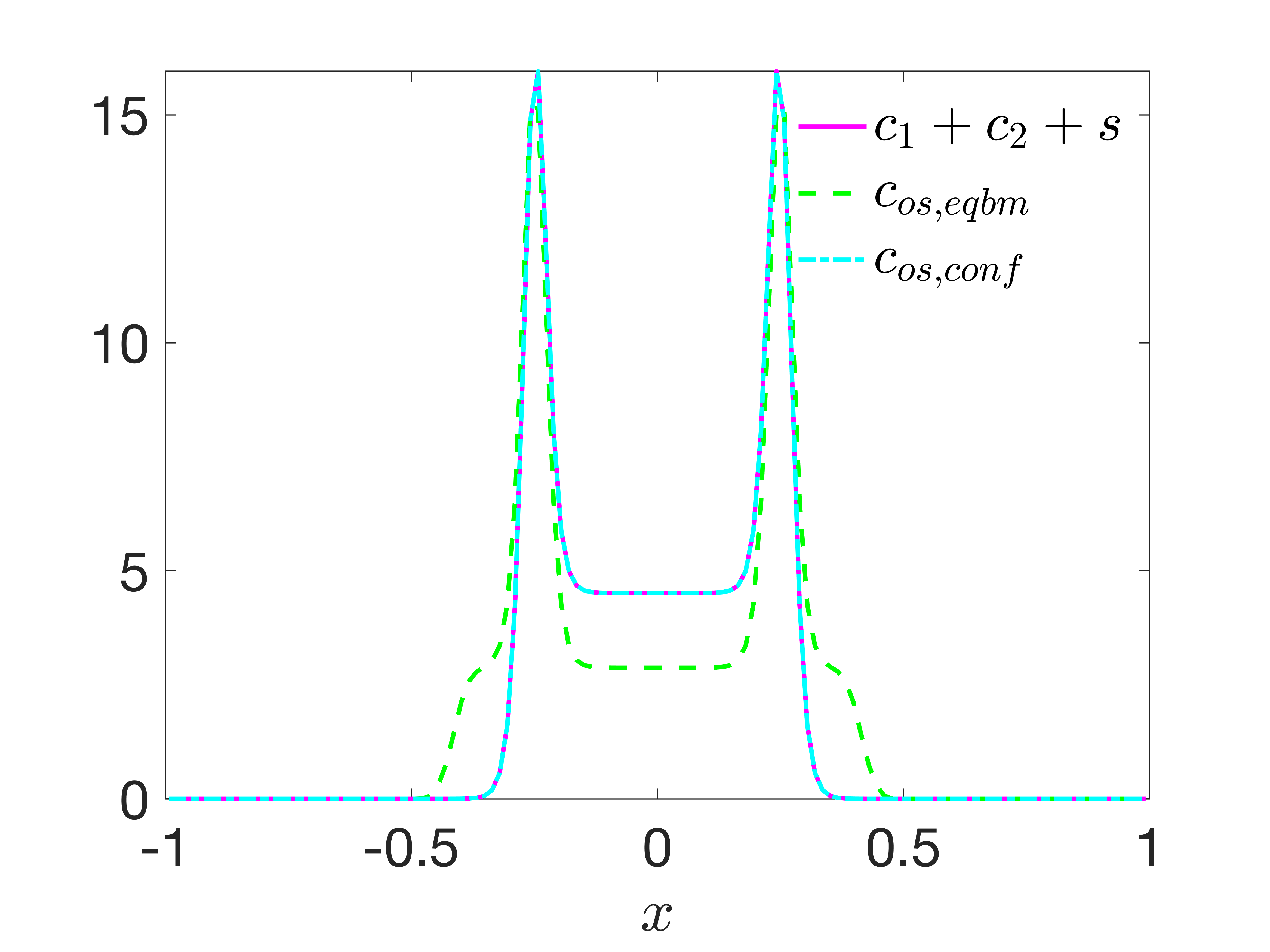}
        \caption{}
    \end{subfigure}
    \begin{subfigure}[b]{0.32\textwidth}
        \includegraphics[width=\textwidth]{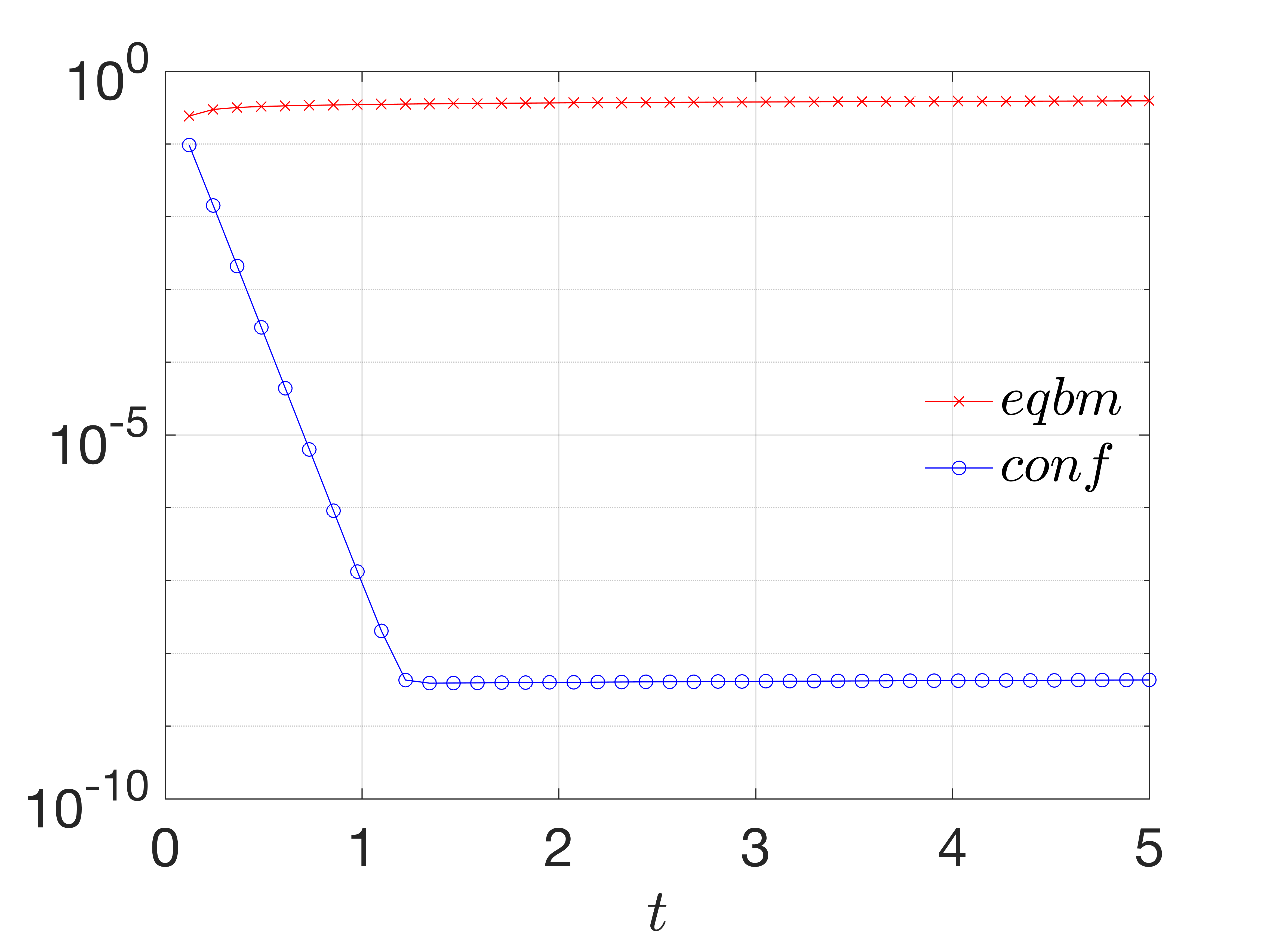}
        \caption{}
    \end{subfigure}

    \caption{One-dimensional transport for a surfactant soluble only in phase 1 (parameters in Table~\ref{tab3}): (a) profiles of $c_1$, $s$, and $\phi$ at multiple time instants from the three-scalar model, (b) total concentration $c_1+c_2+s$ at $t=5$ compared against the full-equilibrium one-scalar model $c_\text{os,eqbm}$ and the confined one-scalar model $c_\text{os,conf}$, and (c) relative $L^1$ differences between the three-scalar model and each one-scalar model as functions of time.}
    \label{fig:1d_case3}
\end{figure}

The simulation parameters are listed in Table~\ref{tab3}. This test covers the case in which the surfactant is insoluble in phase 2, so that $D_2 = r_{a,2} = r_{d,2} = 0$. The system is initialized with $c_1(x,0) = 2.0\phi$, $c_2(x,0) = 0$, and $s(x,0) = 1.5\,\delta$, and integrated to $t = 5$. Since phase 2 plays no role in the surfactant dynamics, the confined one-scalar model is the appropriate equilibrium reduction for this scenario: if an equilibrium assumption is to be made, it must be restricted to phase 1 and the interface. Figure~\ref{fig:1d_case3}(a) shows the time evolution of $c_1$ and $s$ as the system evolves: the initial interfacial concentration gradually equilibrates with the bulk of phase 1 through adsorption and desorption. Panel (b) compares the total $c_1+c_2+s$ at $t = 5$ from the three-scalar model against both one-scalar models. The full-equilibrium model $c_\text{os,eqbm}$ substantially over-predicts the surfactant concentration, as it incorrectly enforces equilibrium with phase 2, which has no transport activity. The three-scalar result is in close agreement with the confined model $c_\text{os,conf}$. Panel (c) confirms this quantitatively: the $L^1$ difference between the three-scalar and confined models decays with time as equilibrium is approached, while that between the three-scalar and full-equilibrium models remains large. Notably, the agreement between the three-scalar and confined models is better here than in Section~\ref{sec:1d_case2}, where the confined assumption was only an approximation (phase 2 was active but slow); here, with $D_2 = r_{a,2} = r_{d,2} = 0$, the confined assumption is exactly satisfied by construction.

\subsubsection{Artificial interface in a single phase system}
\label{sec:1d_case5}
\begin{table}[h!]
\centering
\renewcommand{\arraystretch}{1.3}
\begin{tabular}{|>{\raggedright\arraybackslash}p{4cm} 
                |>{\raggedright\arraybackslash}p{3.5cm} 
                |>{\raggedright\arraybackslash}p{4cm} 
                |>{\raggedright\arraybackslash}p{3.5cm}|}
\hline
 & \textbf{Phase 1} & \textbf{Phase 2} & \textbf{Interface} \\ \hline
\textbf{Initial conditions (at $t = 0$)}
& $\tilde{c}_1 = 5e^{-4x^2}$
& $\tilde{c}_2 = 5e^{-4x^2}$
& $\tilde{s} = 0$ \\ \hline

\textbf{Surfactants parameters}
& $D_1 = 1.0$, $r_{a1} = 0$, $r_{d1} = 0$
& $D_2 = 1.0$, $r_{a2} = 0$, $\textcolor{blue}{\boldsymbol{r_{d2}}} = 0$
& $D_s = 0$, $\tilde{s}_{\infty} = 0$ \\ \hline
\end{tabular}
\caption{Simulation parameters for the test in Section~\ref{sec:1d_case5}. No adsorption or desorption occurs ($r_{a,p} = r_{d,p} = 0$) and the interface is passive ($D_s = s_\infty = 0$).}
\label{tab5}
\end{table}

\begin{figure}[H]
    \captionsetup{labelfont=bf,font=footnotesize}
    \centering

    \begin{subfigure}[b]{0.32\textwidth}
        \includegraphics[width=\textwidth]{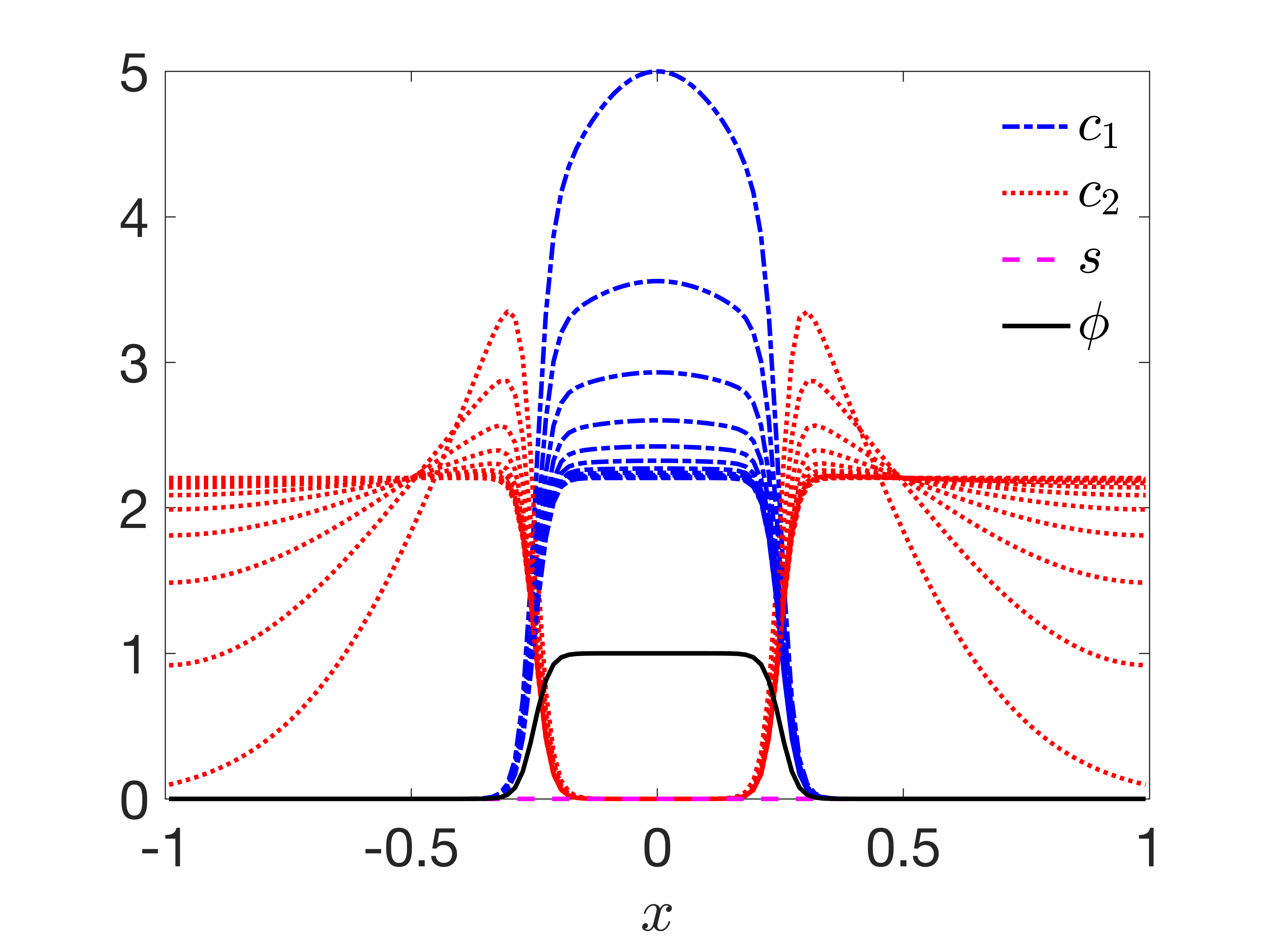}
        \caption{}
    \end{subfigure}
    \hfill
    \begin{subfigure}[b]{0.32\textwidth}
        \includegraphics[width=\textwidth]{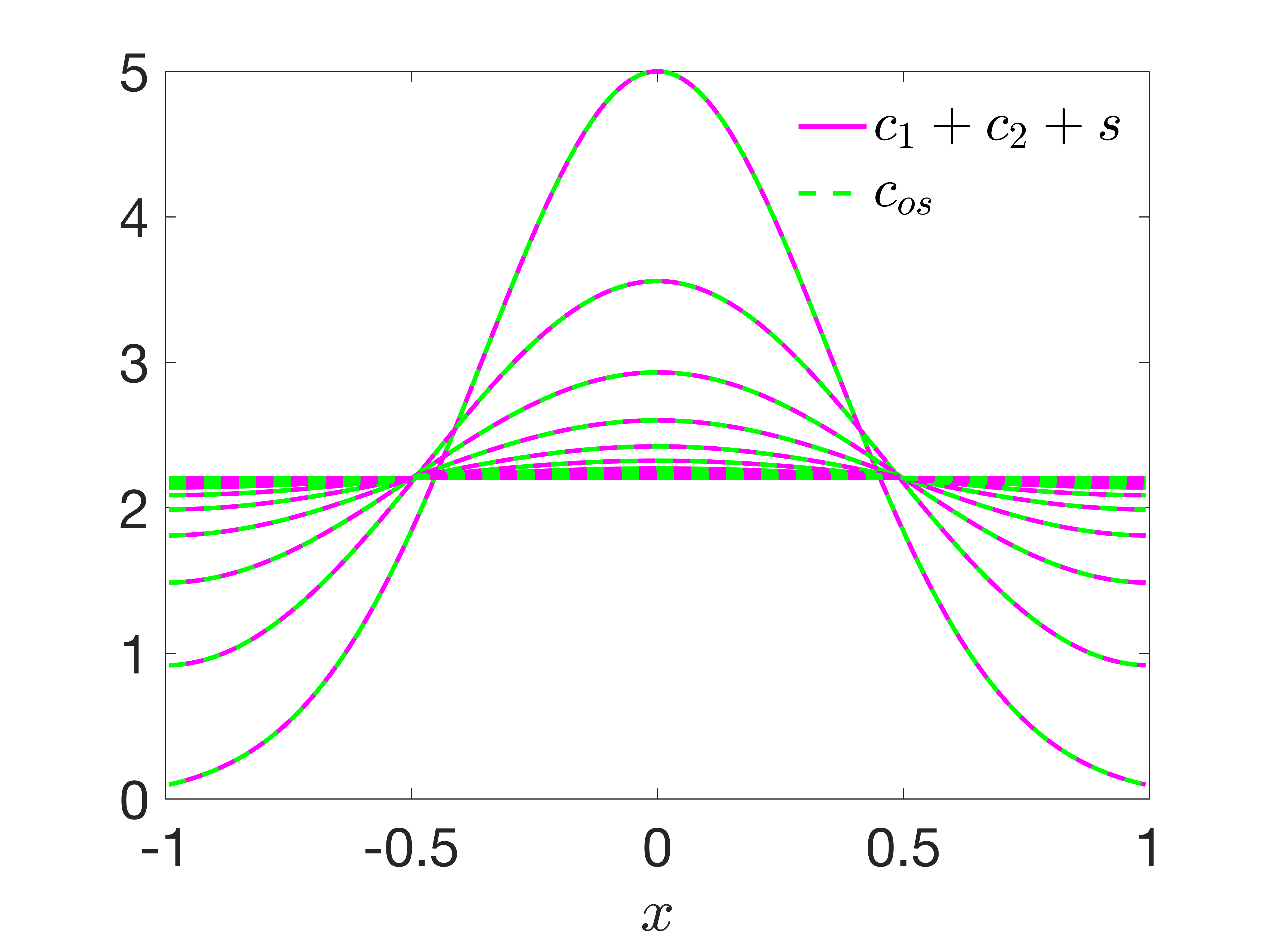}
        \caption{}
    \end{subfigure}
    \begin{subfigure}[b]{0.32\textwidth}
        \includegraphics[width=\textwidth]{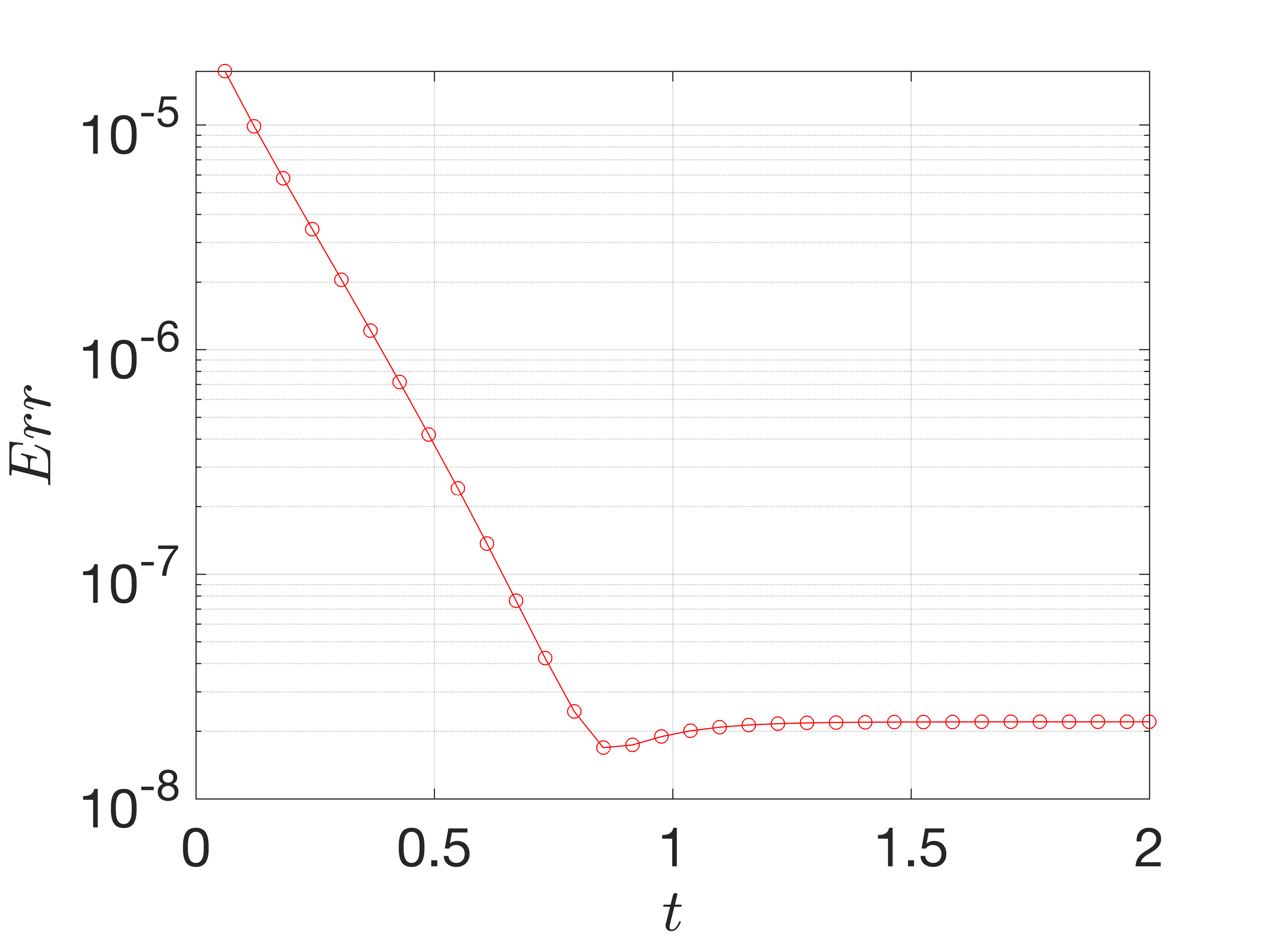}
        \caption{}
    \end{subfigure}

    \caption{One-dimensional transport in the artificial interface limit (parameters in Table~\ref{tab5}): (a) profiles of $c_1$, $c_2$, and $\phi$ at multiple time instants from the three-scalar model, (b) total concentration $c_1+c_2+s$ compared against the equivalent single-phase one-scalar result $c_\text{os}$ at multiple times, and (c) relative $L^1$ error between the three-scalar and one-scalar models as a function of time.}
    \label{fig:1d_case5}
\end{figure}

The simulation parameters are listed in Table~\ref{tab5}. This test verifies the reduction consistency of the three-scalar model in the limit of an artificial interface embedded in a single-phase flow. With $r_{a,p} = r_{d,p} = D_s = s_\infty = 0$, the interface plays no physical role: surfactant neither adsorbs to nor desorbs from it, and $s$ remains identically zero. The model should therefore closely recover single-phase scalar diffusion, with the combined field $c_1 + c_2$ evolving according to a standard diffusion equation. The system is initialized with a Gaussian profile split across the phase field: $c_1(x,0) = 5e^{-4x^2}\phi$ and $c_2(x,0) = 5e^{-4x^2}(1-\phi)$, so that the total $c_1 + c_2 = 5e^{-4x^2}$ is a smooth Gaussian. The simulation is run to $t = 2$. Figure~\ref{fig:1d_case5}(a) shows the evolution of $c_1$ and $c_2$ in the three-scalar model: each field diffuses within its respective domain and the sum $c_1+c_2$ spreads as a standard diffusion profile. Panel (b) compares $c_1+c_2+s$ from the three-scalar model against the equivalent single-phase one-scalar computation $c_\text{os}$, showing very good agreement. Panel (c) shows the relative $L^1$ difference between the two models as a function of time; it remains small throughout, confirming the reduction consistency of the three-scalar model in the artificial interface limit. The small remaining difference arises because the phase-weighted diffusion terms and correction fluxes $\vec{R}$ act on $c_1$ and $c_2$ individually; when summed, both contribute residuals proportional to $(\tilde{c}_1-\tilde{c}_2)$, multiplied by $\nabla\phi$ and $\vec{R}$, respectively, which are confined to the $O(\epsilon)$ interface region. These residuals therefore vanish both as $\epsilon\to 0$ and at equilibrium, where $\tilde{c}_1=\tilde{c}_2$ since $K_\text{eq}=1$, consistent with the findings of \cite{mirjalili2022computational}.

\subsection{Two-dimensional surfactant transport tests}
\label{sec:2D}
These tests use a circular drop in a periodic square domain, centered at the center of the domain at $t = 0$. As in Section~\ref{sec:uncoupled}, the CAC equation is used with $\epsilon = \Delta x$ and $\gamma = 1$.

\subsubsection{Partially soluble surfactant}
\label{sec:2D_soluble}
\begin{table}[H]
\centering
\renewcommand{\arraystretch}{1.3}
\begin{tabular}{|>{\raggedright\arraybackslash}p{4cm}
                |>{\raggedright\arraybackslash}p{3.5cm}
                |>{\raggedright\arraybackslash}p{4cm}
                |>{\raggedright\arraybackslash}p{3.5cm}|}
\hline
 & \textbf{Phase 1} & \textbf{Phase 2} & \textbf{Interface} \\ \hline

\textbf{Initial conditions (at $t = 0$)}
& $\tilde{c}_1 = 2.5$
& $\tilde{c}_2 = 0.8$
& $\tilde{s} = 0$ \\ \hline

\textbf{Surfactants parameters}
& $D_1 = 1.0$, $r_{a1} = 3.0$, $r_{d1} = 1.0$
& $D_2 = 1.0$, $r_{a2} = 1.0$, $\textcolor{blue}{\boldsymbol{r_{d2}}} = \frac{1}{3}$
& $D_s = 0.2$, $\tilde{s}_{\infty} = 1.0$ \\ \hline
\end{tabular}
\caption{Simulation parameters for the test in Section~\ref{sec:2D_soluble}. The partition coefficient is $K_\text{eq} = 1$.}
\label{tab:2d_soluble}
\end{table}

\begin{figure}[h!]
    \centering
    \begin{subfigure}[b]{0.48\textwidth}
        \centering
        \includegraphics[width=\textwidth]{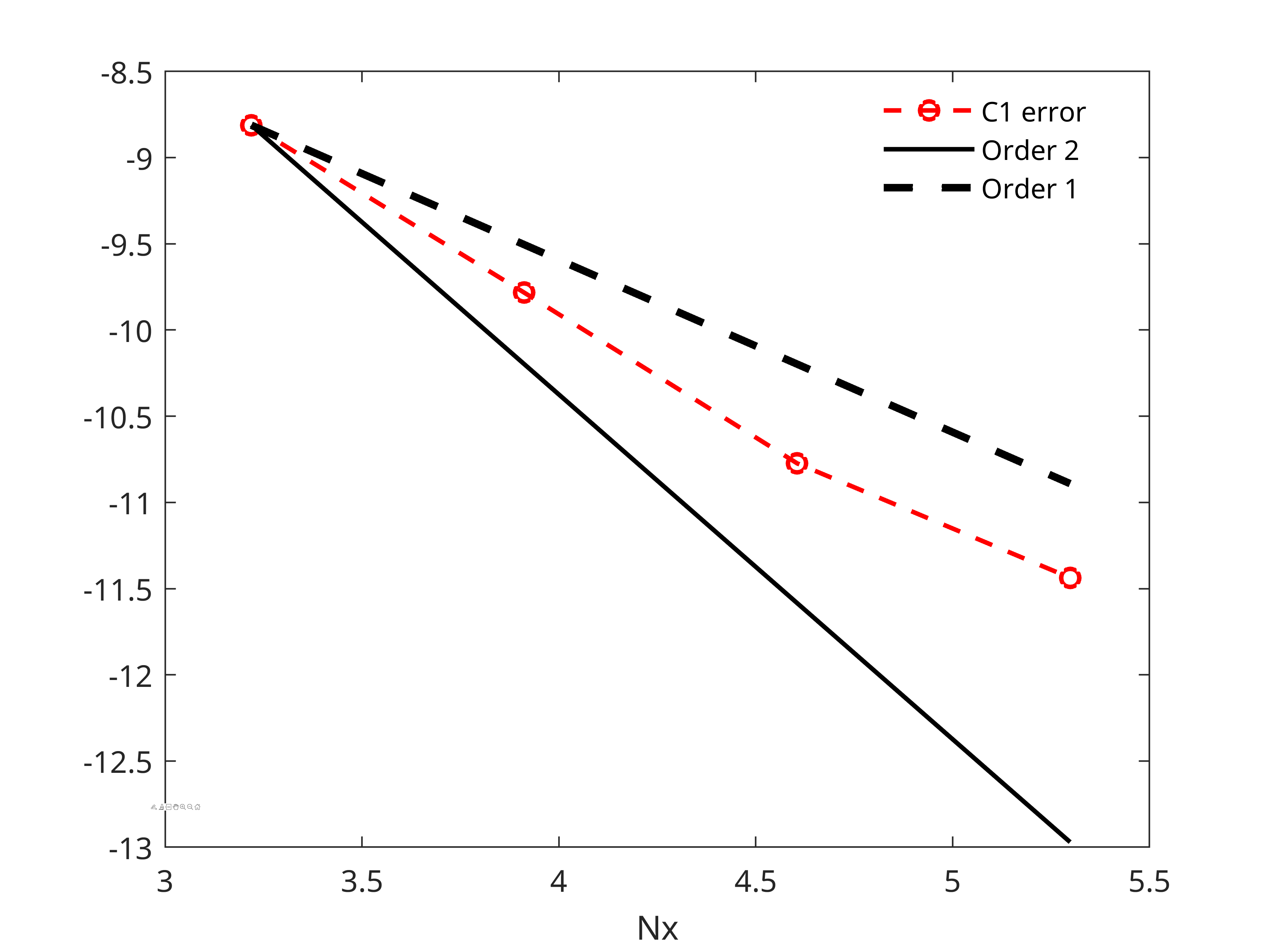}
        \caption{}
        \label{c1_1}
    \end{subfigure}
    \hfill
    \begin{subfigure}[b]{0.48\textwidth}
        \centering
        \includegraphics[width=\textwidth]{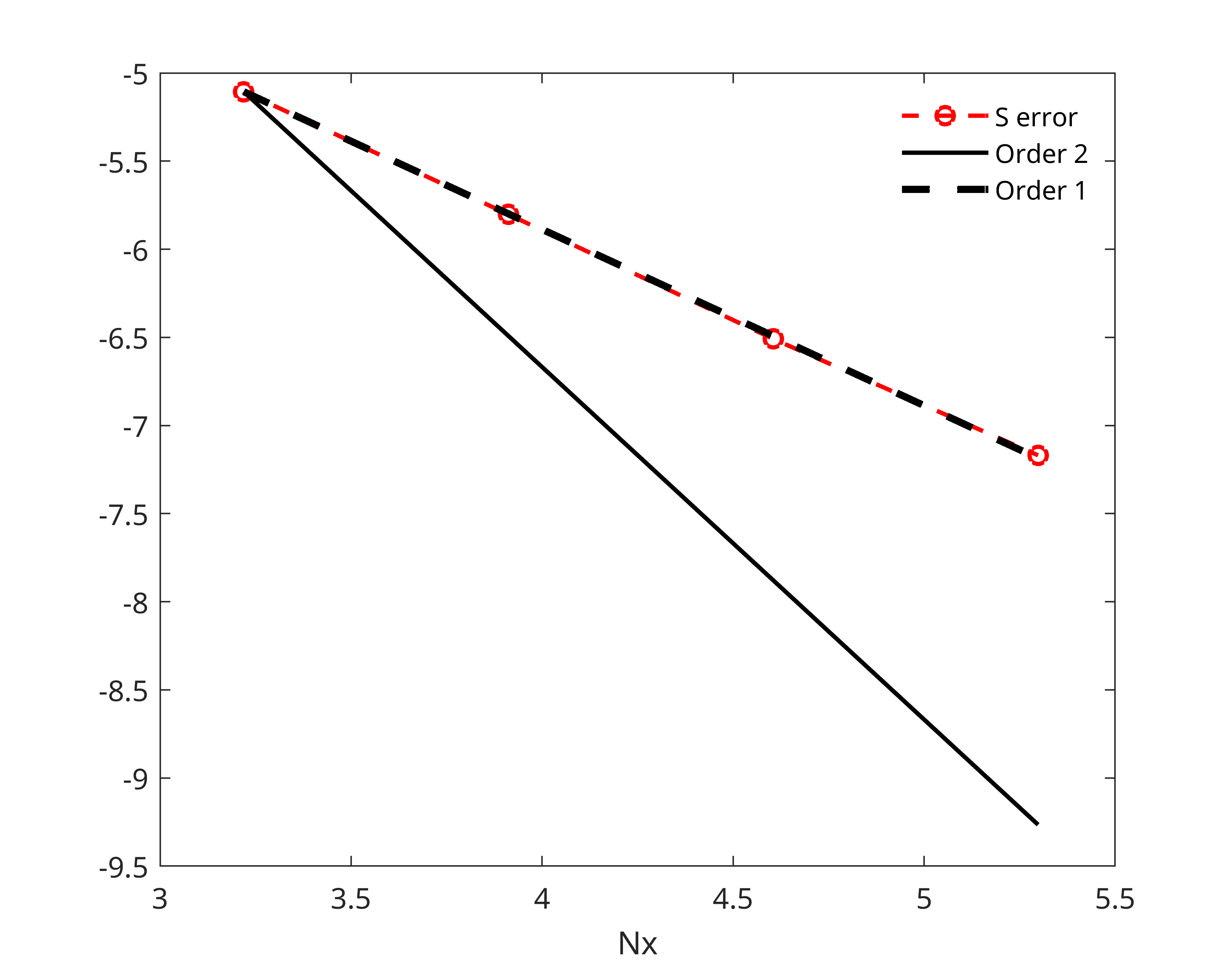}
        \caption{}
        \label{S_1}
    \end{subfigure}
    \caption{Relative $L^2$ errors as functions of grid spacing at $t = 2$ for the two-dimensional partially soluble test (Section~\ref{sec:2D_soluble}): (a) error in $c_1$ and (b) error in $s$. Dashed lines indicate first- and second-order reference slopes.}
    \label{figure_orders_case1}
\end{figure}

\begin{figure}[H]
    \centering
    \includegraphics[width=0.4\textwidth]{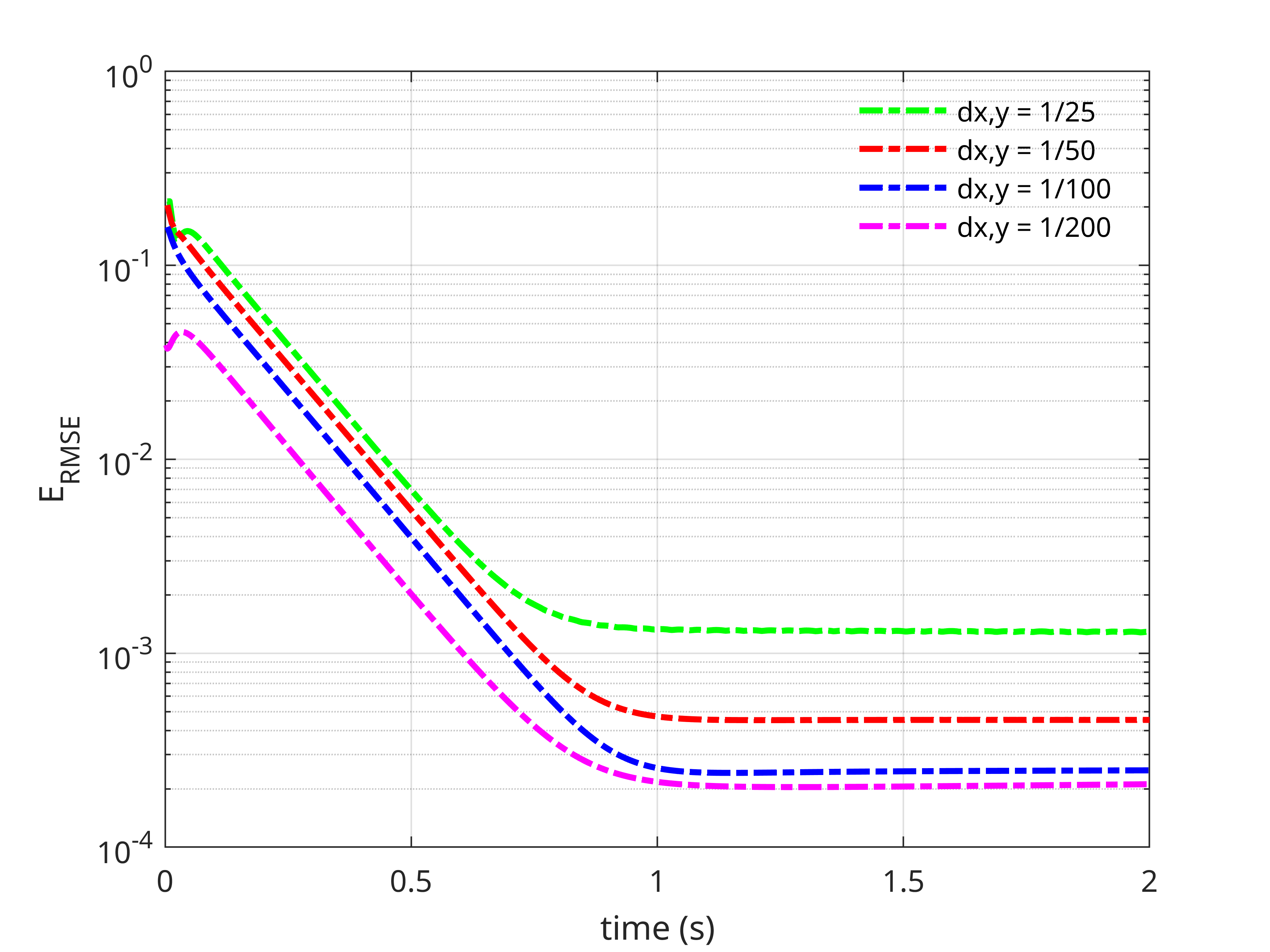}
    \caption{Root-mean-square difference $E_\text{RMSE}$ (Eq.~\eqref{eqn:RMSE}) between the three-scalar and one-scalar model predictions as a function of grid spacing at $t = 2$ (Section~\ref{sec:2D_soluble}).}
    \label{meshref}
\end{figure}

\begin{figure}[H]
    \centering
    \begin{subfigure}{0.48\textwidth}
        \includegraphics[width=\linewidth]{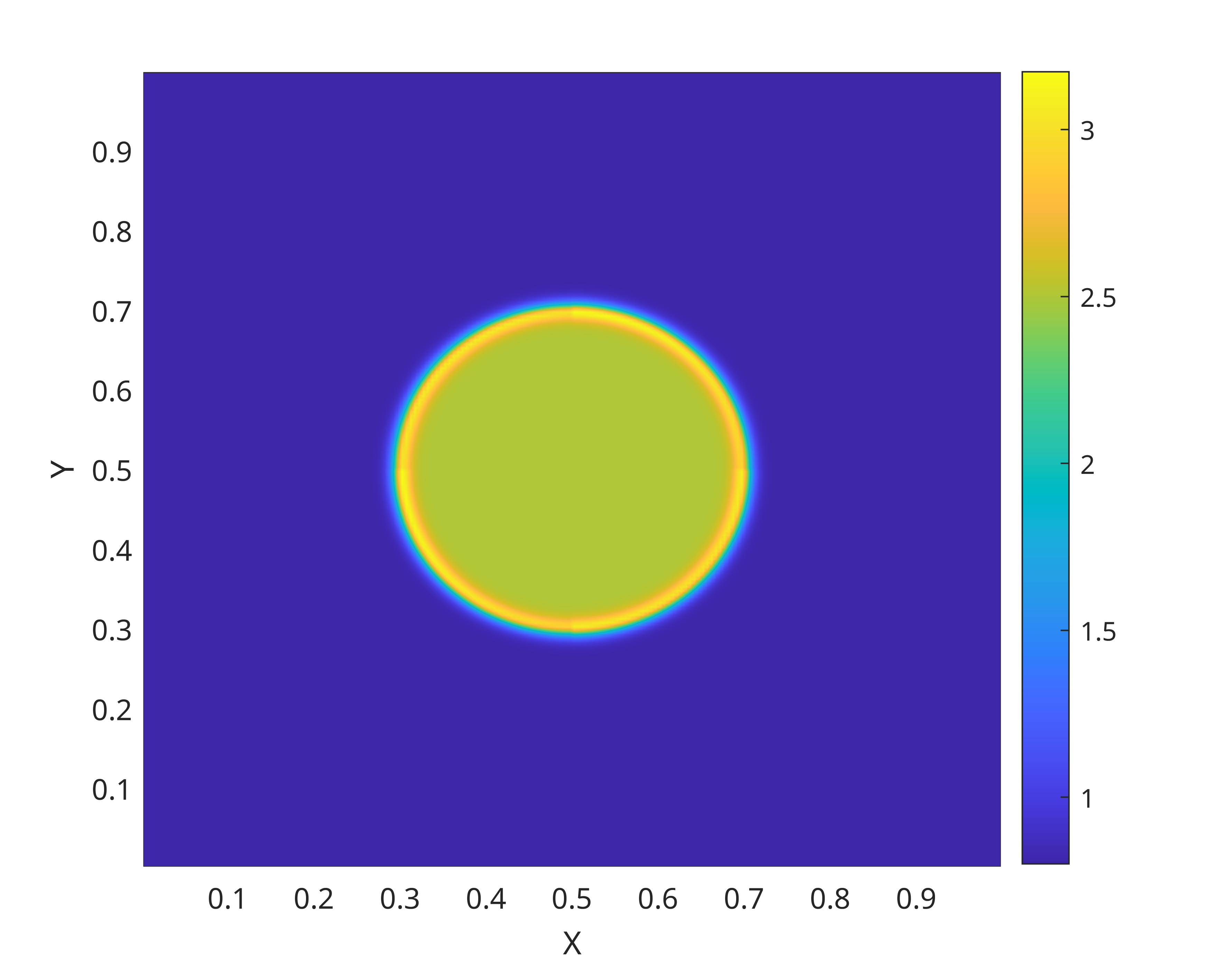}
        \caption{t = 0 s}
    \end{subfigure}
    \hfill
    \begin{subfigure}{0.48\textwidth}
        \includegraphics[width=\linewidth]{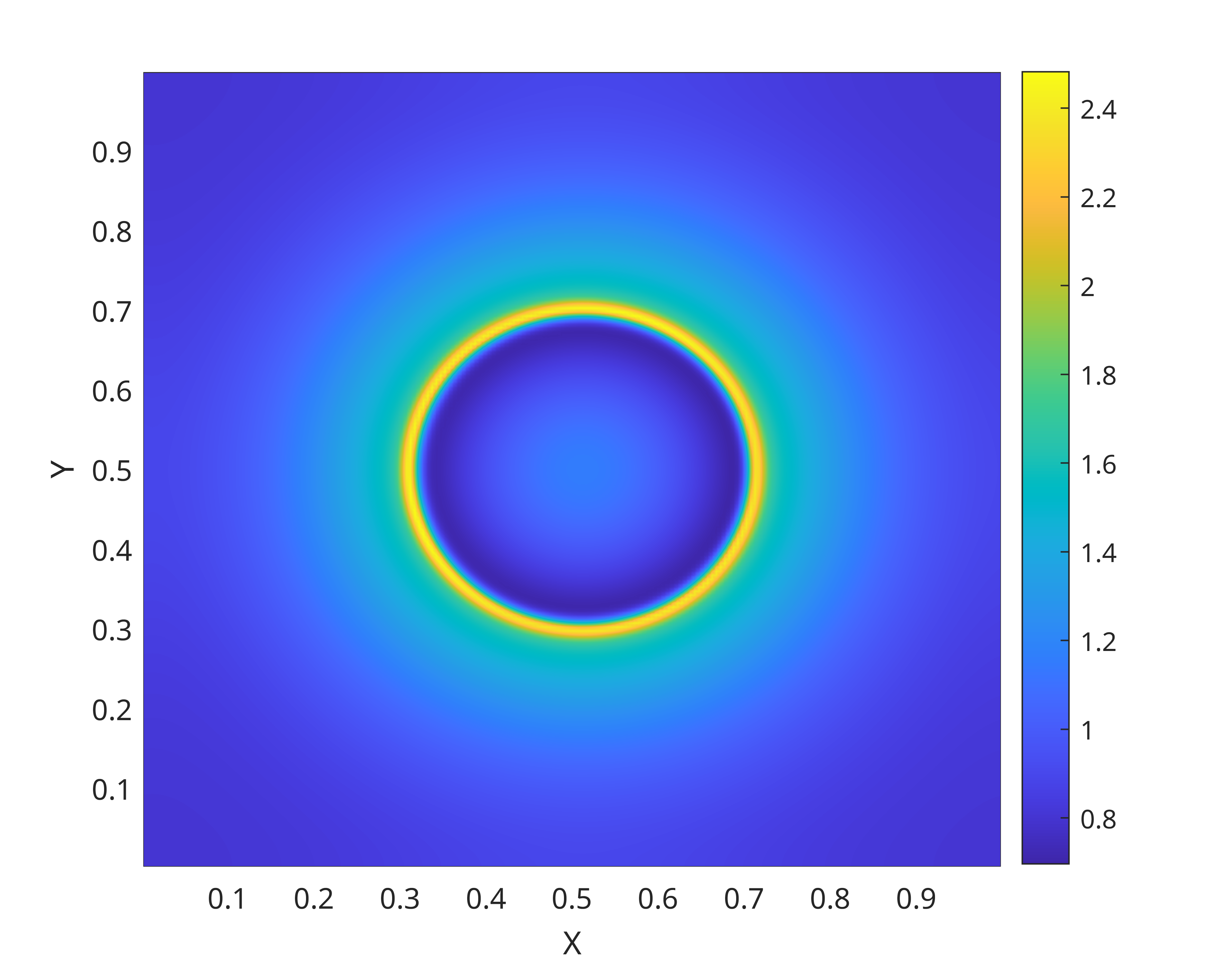}
        \caption{t = 0.0125 s}
    \end{subfigure}

    \begin{subfigure}{0.48\textwidth}
        \includegraphics[width=\linewidth]{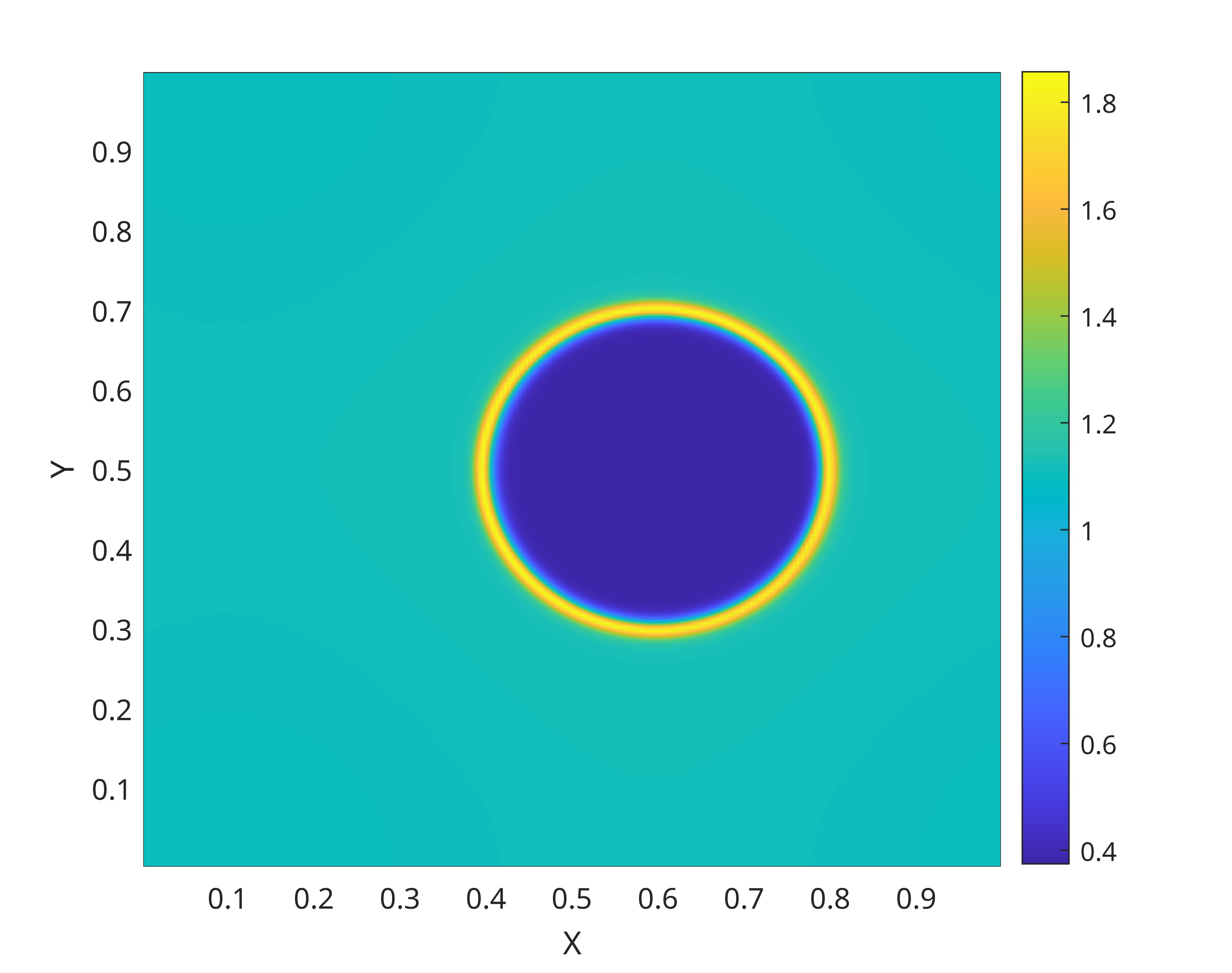}
        \caption{t = 0.1 s}
    \end{subfigure}
    \hfill
    \begin{subfigure}{0.48\textwidth}
        \includegraphics[width=\linewidth]{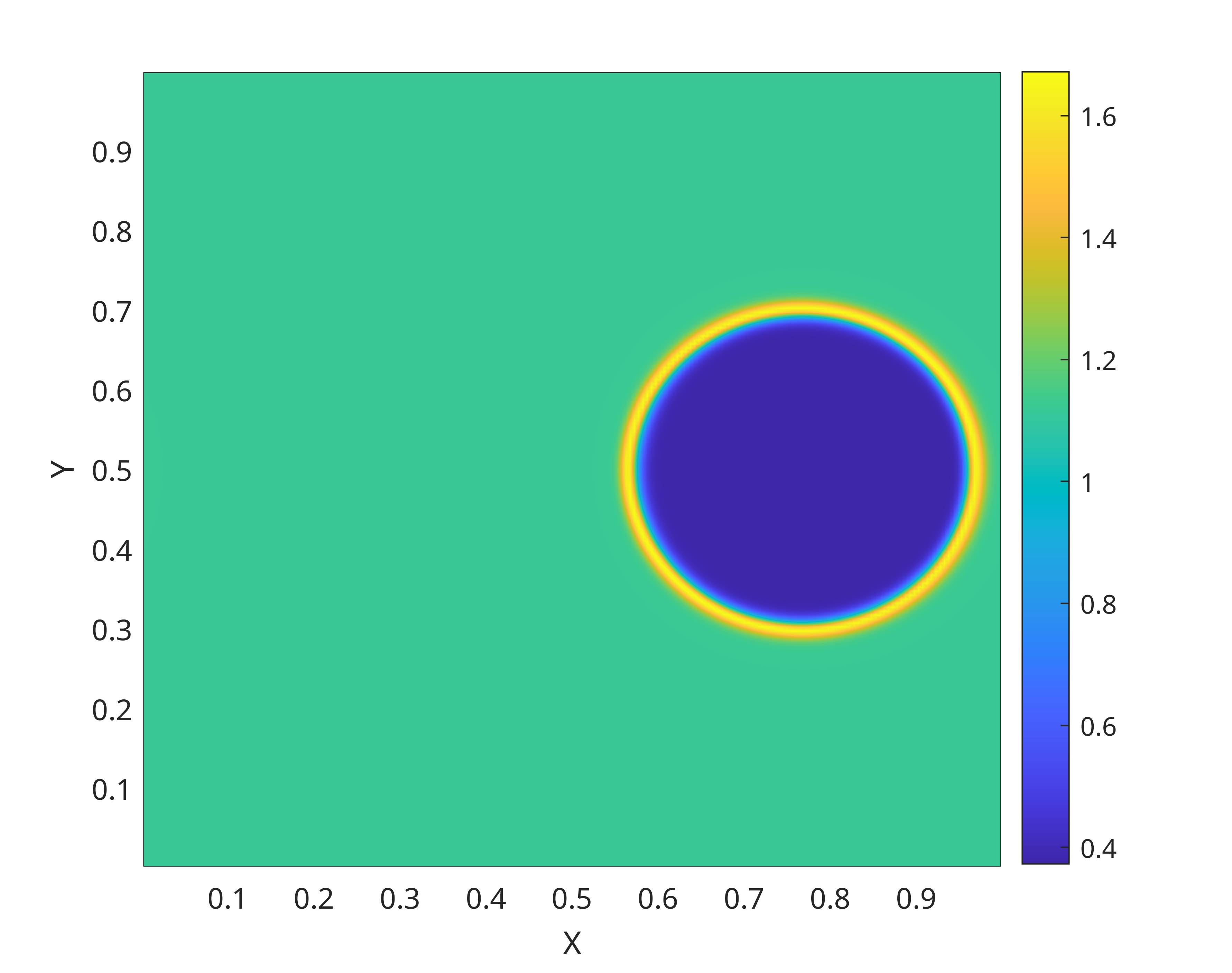}
        \caption{t = 0.26 s }
    \end{subfigure}

    \begin{subfigure}{0.48\textwidth}
        \includegraphics[width=\linewidth]{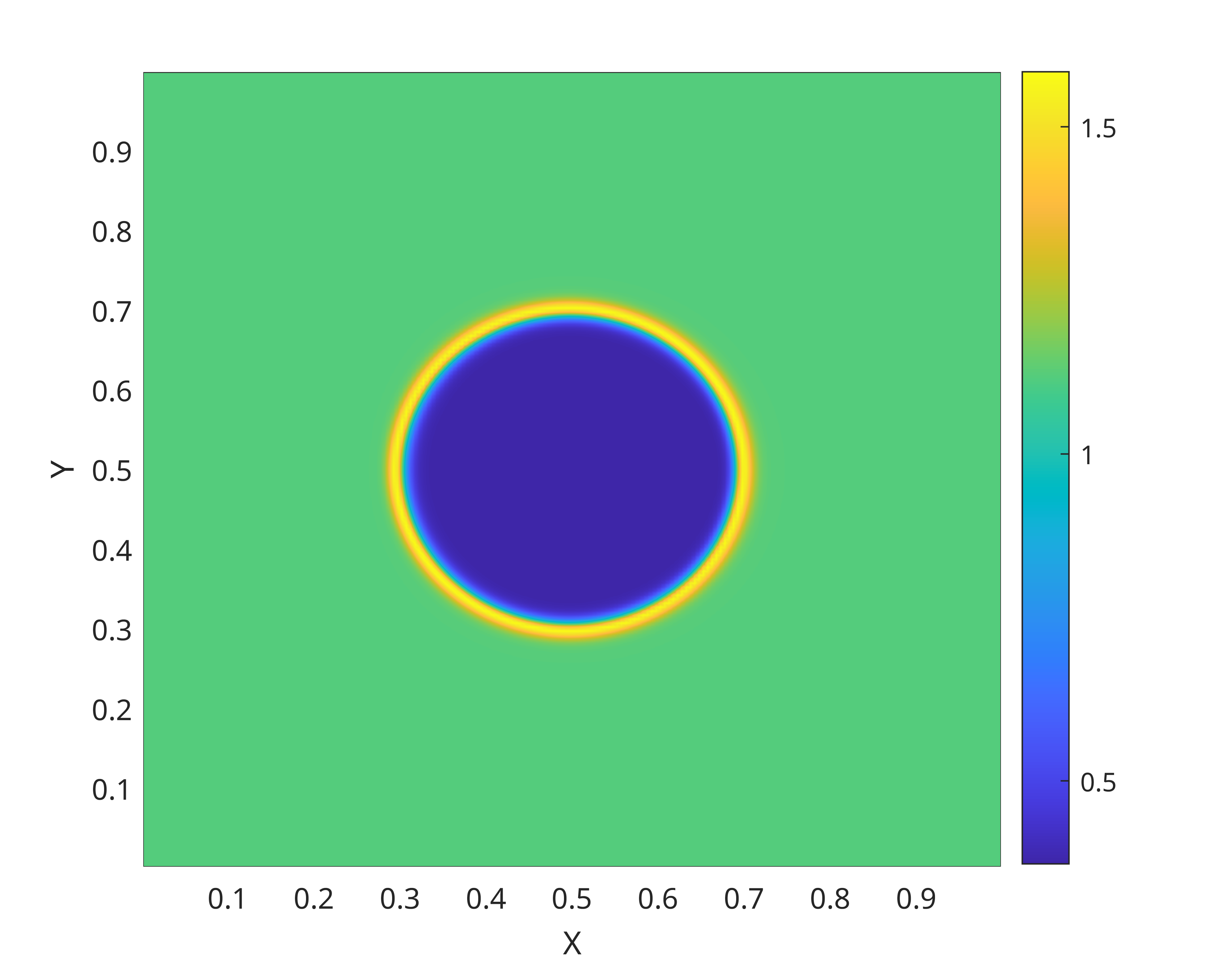}
        \caption{t = 1 s}
    \end{subfigure}
    \hfill
    \begin{subfigure}{0.48\textwidth}
        \includegraphics[width=\linewidth]{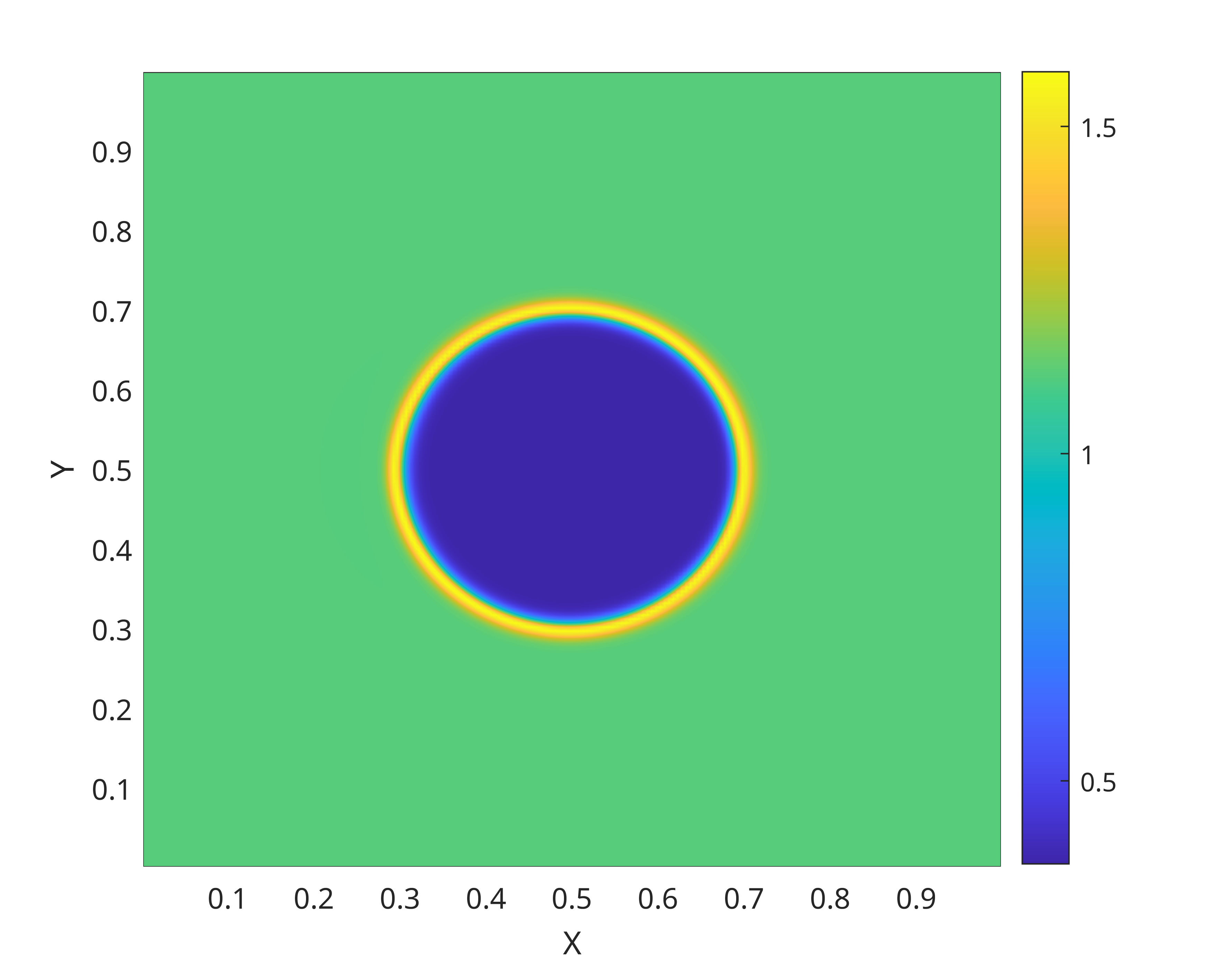}
        \caption{t = 2 s}
    \end{subfigure}

    \caption{Evolution of the total concentration $c_1 + c_2 + s$ at times $t = 0$, $0.0125$, $0.0975$, $0.26$, $1$, and $2$ for the two-dimensional partially soluble test (parameters in Table~\ref{tab:2d_soluble}).}
    \label{evolution1}
\end{figure}

The simulation parameters are listed in Table~\ref{tab:2d_soluble}, which match those of the one-dimensional test in Section~\ref{sec:1d_case1}. The drop has radius $R = 0.2$ in a $[0,1]^2$ periodic domain. The drop is advected with uniform velocity $\vec{u} = \hat{e}_x$ while diffusion, adsorption, and desorption simultaneously drive the system toward the Langmuir equilibrium. The accuracy of the numerical solution is quantified at $t = 2$ by the relative $L^2$ errors of $c_1$ and $s$ with respect to their equilibrium values $c_{1,\text{eq}}$ and $s_\text{eq}$. These equilibrium references are evaluated locally using the numerical $\tilde{c}_2 = c_2/\phi_2$ field at $t = 2$: $s_\text{eq}$ is obtained from the Langmuir isotherm (Eq.~\eqref{eqn:surf_transfer} with $J_{ps}=0$, as in Section~\ref{sec:1d_case1}) and $c_{1,\text{eq}} = K_\text{eq}\,\tilde{c}_2\,\phi_1$. The errors are defined as
\begin{equation}
E_{L^2}(q) = \frac{\left\| q - q_\text{eq} \right\|_2}{\left\| q_\text{eq} \right\|_2}, \quad q \in \{c_1,\, s\}.
\label{eqn:L2_error}
\end{equation}
Table~\ref{error_C1_S_1} reports the errors at four grid spacings, and Figure~\ref{figure_orders_case1} plots them as a function of $\Delta x$. Both $c_1$ and $s$ show approximately first-order convergence with mesh refinement, consistent with the $\epsilon = \Delta x$ scaling of the interface thickness. Figure~\ref{meshref} plots the root-mean-square difference between the three-scalar and one-scalar model predictions as the mesh is refined, defined as
\begin{equation}
E_\text{RMSE} = \sqrt{\frac{1}{N}\sum_{i=1}^{N}\left[(c_1+c_2+s)^i - c_\text{os}^i\right]^2};
\label{eqn:RMSE}
\end{equation}
the two models converge to the same solution as the mesh is refined. Figure~\ref{evolution1} shows the evolution of the total concentration $c_1 + c_2 + s$ at representative time instants.

\begin{table}[h!]
    \centering
    \setlength{\tabcolsep}{10pt}
    \renewcommand{\arraystretch}{1.2}
    \begin{tabular}{|c|c|c|}
        \hline
        \textbf{Grid spacing} & \textbf{$c_1$ error} ($\times 10^{-5}$) & \textbf{$s$ error} ($\times 10^{-3}$) \\
        \hline
        $1/25$  & 14.9 & 6.06\\
        $1/50$  &  5.64 & 3.02 \\
        $1/100$ &  2.09 & 1.49 \\
        $1/200$ &  1.08 & 0.77 \\
        \hline
    \end{tabular}
    \caption{Relative $L^2$ errors for $c_1$ and $s$ at $t = 2$ for the test in Section~\ref{sec:2D_soluble}.}
    \label{error_C1_S_1}
\end{table}

\subsubsection{Insoluble surfactant}
\label{sec:2d_insoluble}
\begin{table}[H]
\centering
\renewcommand{\arraystretch}{1.3}
\begin{tabular}{|>{\raggedright\arraybackslash}p{4cm}
                |>{\raggedright\arraybackslash}p{3.5cm}
                |>{\raggedright\arraybackslash}p{4cm}
                |>{\raggedright\arraybackslash}p{3.5cm}|}
\hline
 & \textbf{Phase 1} & \textbf{Phase 2} & \textbf{Interface} \\ \hline

\textbf{Initial conditions (at $t = 0$)}
& $\tilde{c}_1 = 0$
& $\tilde{c}_2 = 0$
& $\tilde{s}_0(\theta)$ \\ \hline

\textbf{Surfactants parameters}
& $D_1 = 0$, $r_{a1} = 0$, $r_{d1} = 0$
& $D_2 = 0$, $r_{a2} = 0$, $r_{d2} = 0$
& $D_s = 1.0$, $\tilde{s}_{\infty} = 1.0$ \\ \hline
\end{tabular}
\caption{Simulation parameters for the test in Section~\ref{sec:2d_insoluble}. The surfactant is insoluble ($D_p = r_{a,p} = r_{d,p} = 0$).}
\label{tab:2d_insoluble}
\end{table}

\begin{figure}[h!]
\captionsetup{labelfont=bf,font=footnotesize}
\centering
\begin{minipage}{0.65\textwidth}
    \begin{subfigure}[b]{\textwidth}
        \includegraphics[width=\textwidth]{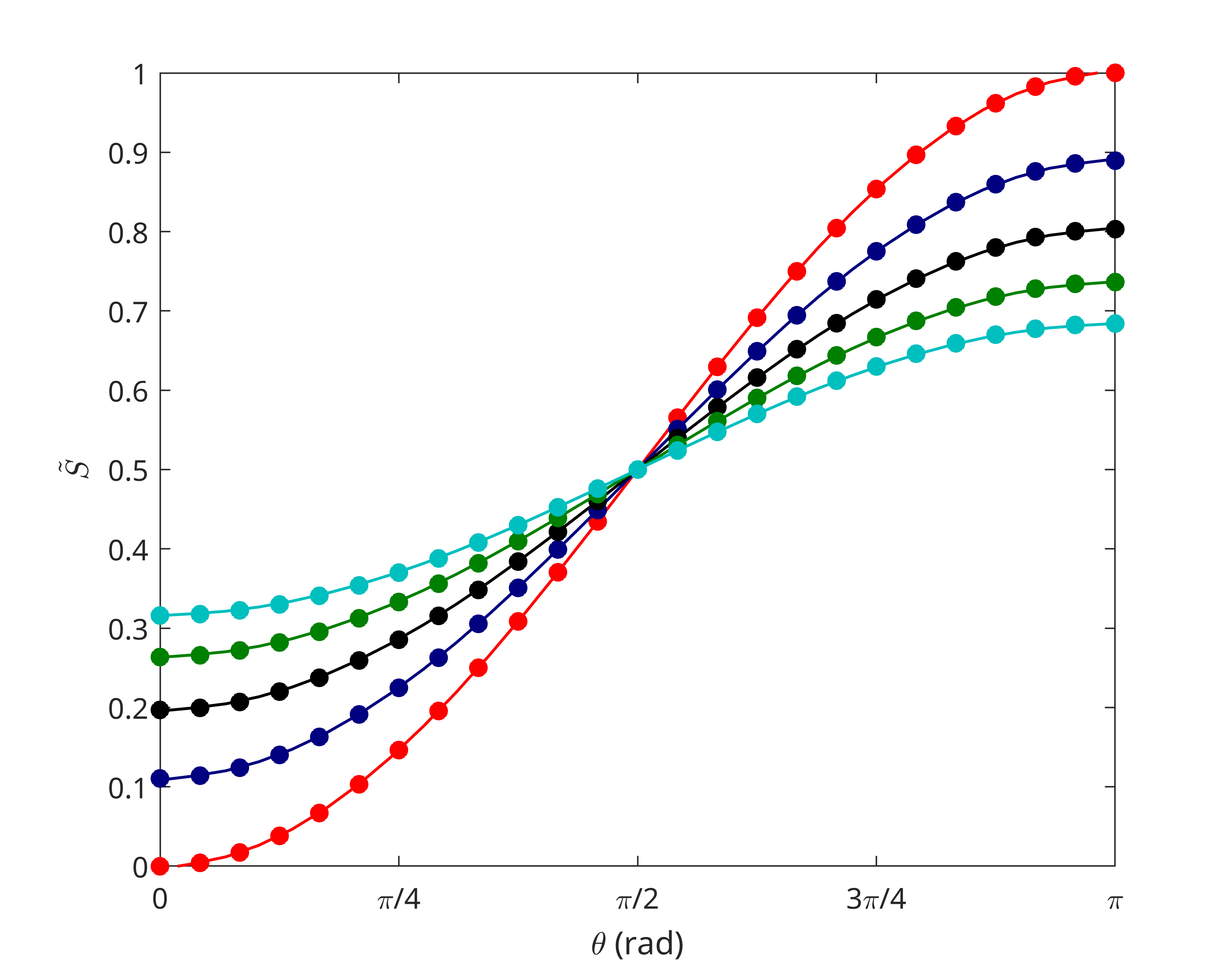}
        \caption{$\Delta x = 1/100$}
    \end{subfigure}
    \vspace{0.5em}
    \begin{subfigure}[b]{\textwidth}
        \includegraphics[width=\textwidth]{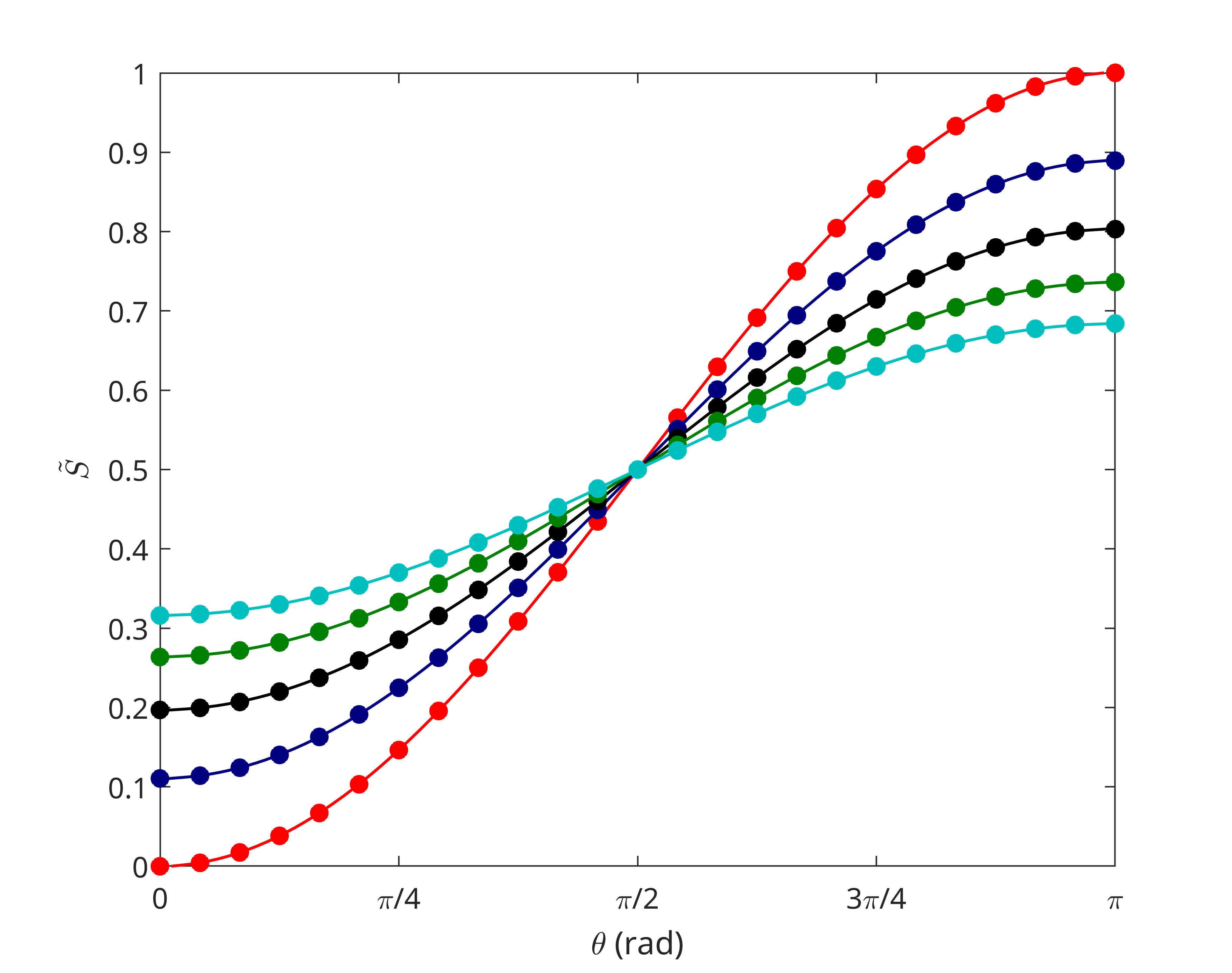}
        \caption{$\Delta x = 1/200$}
    \end{subfigure}
\end{minipage}
\hfill
\begin{minipage}{0.3\textwidth}
    \centering
    \textbf{Legend:} \\
    \begin{tabular}{l}
    {\color{red} --- Numerical $t=0$} \\
    {\color{red} \textbullet\ Analytical $t=0$} \\
    {\color{darkblue} --- Numerical $t=0.25$} \\
    {\color{darkblue} \textbullet\ Analytical $t=0.25$} \\
    {\color{black} --- Numerical $t=0.5$} \\
    {\color{black} \textbullet\ Analytical $t=0.5$} \\
    {\color{darkgreen} --- Numerical $t=0.75$} \\
    {\color{darkgreen} \textbullet\ Analytical $t=0.75$} \\
    {\color{turquoise} --- Numerical $t=1$} \\
    {\color{turquoise} \textbullet\ Analytical $t=1$} \\
    \end{tabular}
\end{minipage}
\caption{Specific interfacial concentration $\tilde{s}$ as a function of polar angle $\theta$ at five time instants, compared against the analytical solution Eq.~\eqref{eqn:insoluble_ana} (parameters in Table~\ref{tab:2d_insoluble}), at grid spacings (a) $\Delta x = 1/100$ and (b) $\Delta x = 1/200$.}
\label{comparaison_insoluble}
\end{figure}

\begin{figure}
\captionsetup{labelfont=bf,font=footnotesize}
    \centering
    \begin{subfigure}{0.48\textwidth}
        \includegraphics[width=\linewidth]{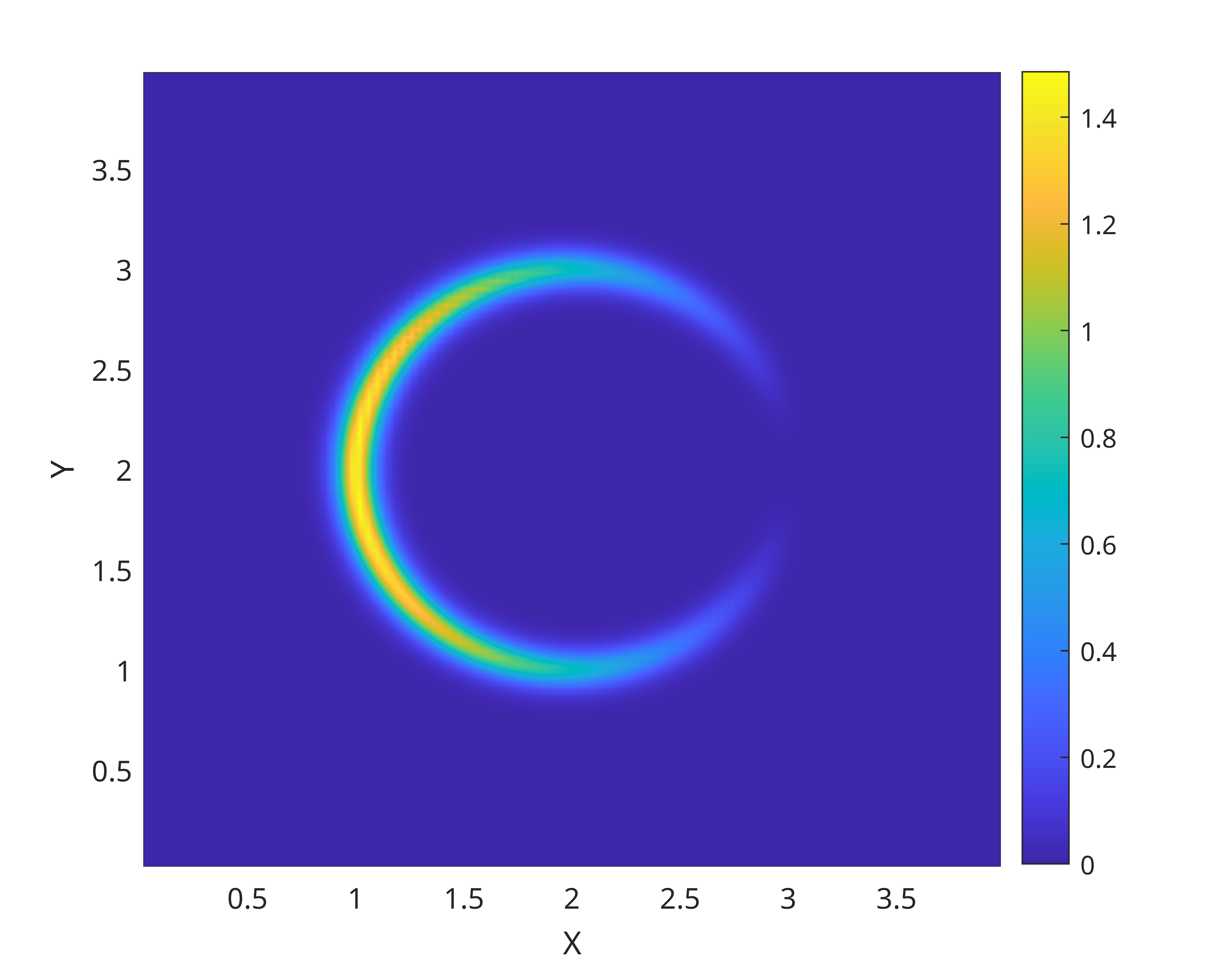}
        \caption{$\Delta x = 1/100$, $t = 0$}
    \end{subfigure}
    \hfill
    \begin{subfigure}{0.48\textwidth}
        \includegraphics[width=\linewidth]{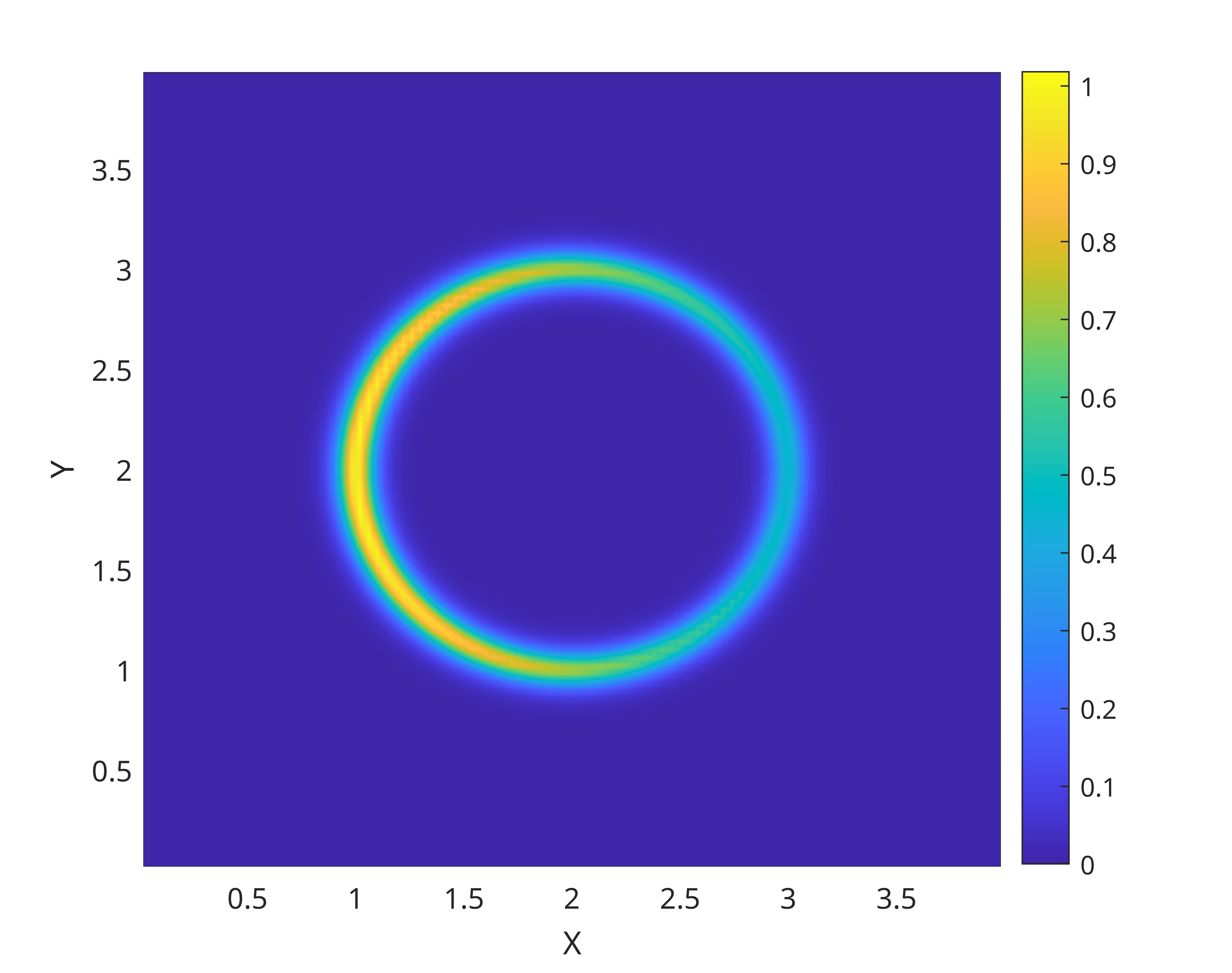}
        \caption{$\Delta x = 1/100$, $t = 1$}
    \end{subfigure}
    \vspace{0.5em}
    \begin{subfigure}{0.48\textwidth}
        \includegraphics[width=\linewidth]{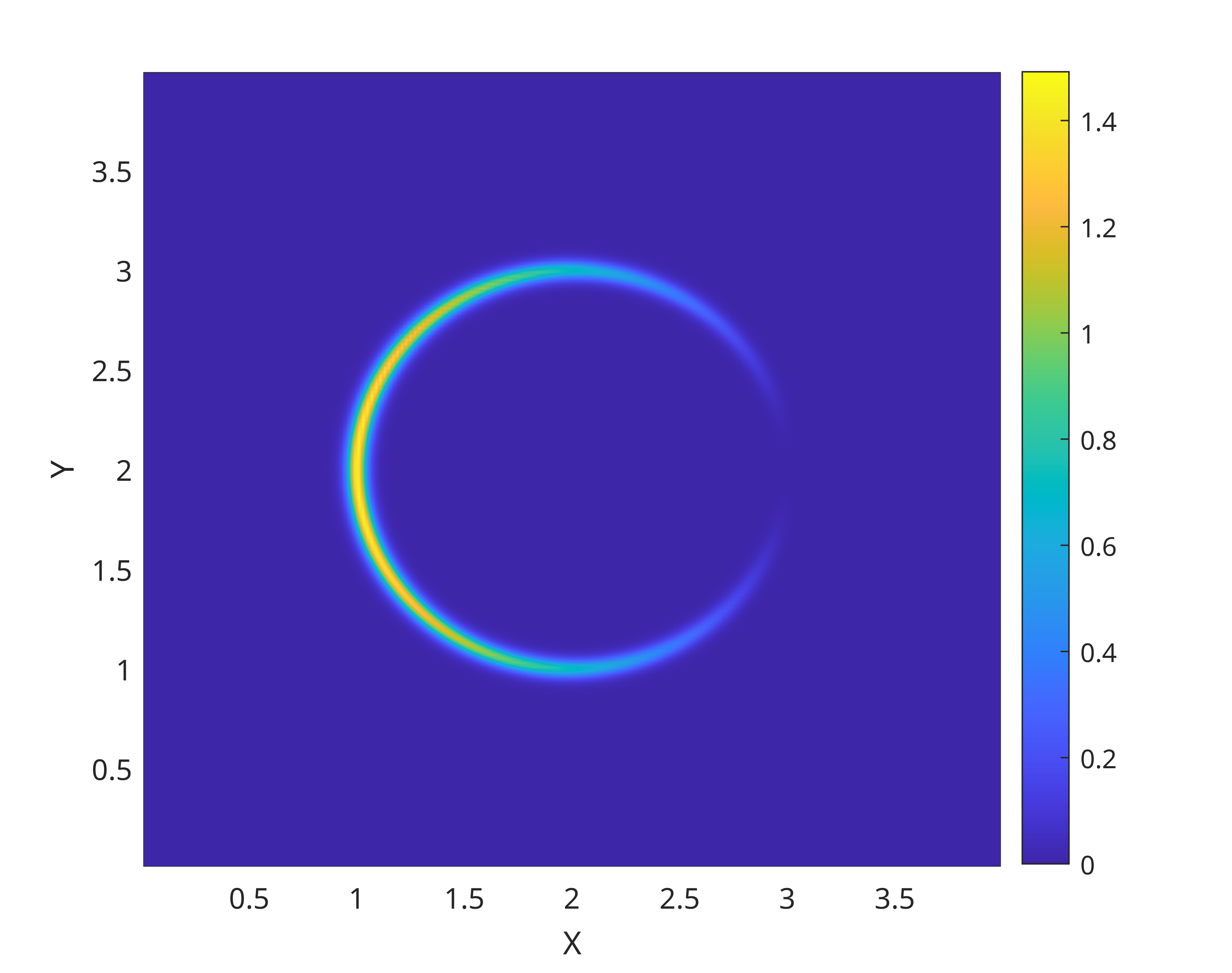}
        \caption{$\Delta x = 1/200$, $t = 0$}
    \end{subfigure}
    \hfill
    \begin{subfigure}{0.48\textwidth}
        \includegraphics[width=\linewidth]{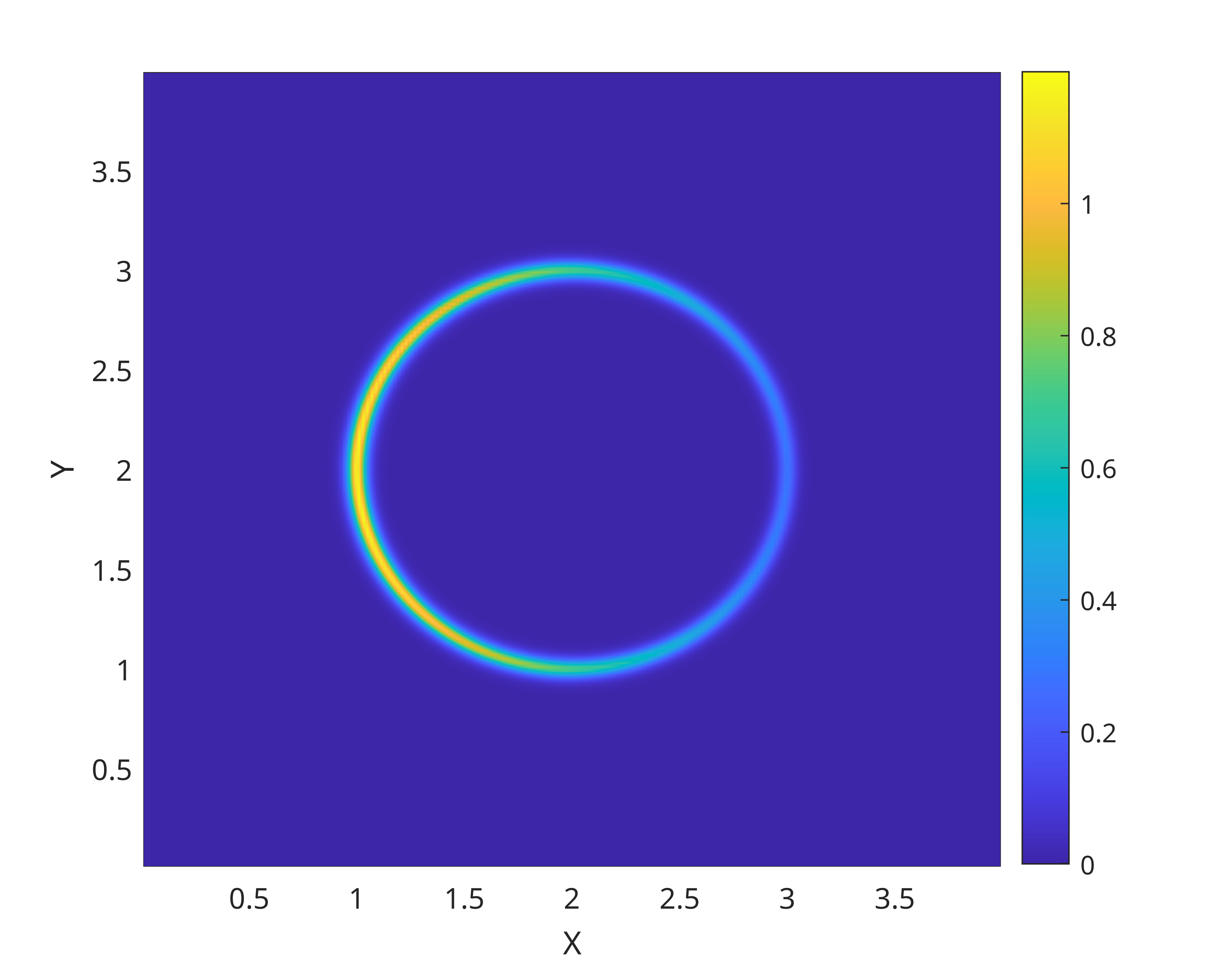}
        \caption{$\Delta x = 1/200$, $t = 1$}
    \end{subfigure}
    \caption{Two-dimensional field of $s$ for the insoluble surfactant test (Section~\ref{sec:2d_insoluble}) at $t = 0$ and $t = 1$, for grid spacings $\Delta x = 1/100$ (a,b) and $\Delta x = 1/200$ (c,d).}
    \label{evolution2}
\end{figure}

The simulation parameters are listed in Table~\ref{tab:2d_insoluble}. This test follows the setup of \cite{jain_surf1}, using a drop of radius $R = 1$ in a $[0,4]^2$ periodic domain. The surfactant is insoluble and confined to the interface, with the initial specific interfacial concentration given by
\begin{equation}
    \tilde{s}_0(\theta) = \frac{1}{2}(1-\cos\theta),
\end{equation}
where $\theta$ is the polar angle measured around the drop. The velocity is zero; the drop is stationary and the surfactant evolves solely by diffusion on the circular interface. The analytical solution is
\begin{equation}
    \tilde{s}_\text{ana}(\theta,t) = \frac{1}{2}\!\left(1 - e^{-D_s t/R^2}\cos\theta\right),
    \label{eqn:insoluble_ana}
\end{equation}
where $R = 1$ is the drop radius, with the $\cos\theta$ mode decaying toward the uniform distribution $\tilde{s} = 1/2$ with time constant $R^2/D_s = 1$ \citep{jain_surf1}. Figure~\ref{comparaison_insoluble} shows $\tilde{s}$ as a function of $\theta$ at times $t = 0$, $0.25$, $0.5$, $0.75$, and $1$, compared against Eq.~\eqref{eqn:insoluble_ana}, for grids of $100\times100$ and $200\times200$. The numerical solution is in close agreement with the analytical prediction at both resolutions and shows improved accuracy compared to the results reported in Fig.~6 of \cite{jain_surf1} at equivalent resolution. Figure~\ref{evolution2} shows the corresponding two-dimensional fields of $s$ at $t = 0$ and $t = 1$ for both resolutions.

\subsection{Flow-coupled tests}
\label{sec:flow_coupled}
These tests follow the setup of Section~5 of \cite{Teigen2011} for a surfactant-laden drop in a two-dimensional shear flow. A circular drop of radius $R = 1$ is placed at the center of a domain $[-6,6]\times[-2,2]$ subject to the background shear $\vec{u} = (\dot{\gamma}y,\,0)$ with $\dot{\gamma} = 1$, imposed through Dirichlet boundary conditions at the top and bottom walls and periodic conditions in the streamwise direction. Both phases have equal density and viscosity; the capillary number is $Ca = \mu\dot{\gamma}/\sigma_0 = 0.5$ and the Reynolds number is $Re = \rho\dot{\gamma}R^2/\mu = 1.0$. As noted in Section~\ref{sec:three_scalar}, these tests use $\delta = 6\phi_1^2\phi_2^2/\epsilon$ to match \cite{Teigen2011}. All simulations use a uniform $384\times128$ mesh with $\epsilon = \Delta x$; no adaptive refinement is required, in contrast to the block-structured adaptive solver of \cite{Teigen2011}. The dependence of surface tension on the specific interfacial concentration $\tilde{s}$ is governed by the Langmuir equation of state \citep{Teigen2011},
\begin{equation}
\sigma(\tilde{s}) = 1 + \beta\ln(1 - \omega\tilde{s}),
\label{eqn:EOS}
\end{equation}
with elasticity number $\beta = 0.3$ and dimensionless surfactant coverage $\omega = \tilde{s}_0/\tilde{s}_\infty = 0.5$, where $\tilde{s}_0 = 1$ is the initial uniform surface concentration and $\tilde{s}_\infty = 1/\omega = 2$ is the saturation concentration. The surface Peclet number $Pe_s = \dot{\gamma}R^2/D_s = 10$ is held fixed. For the soluble tests, the Biot number $Bi = r_{d,p}/\dot{\gamma}$ measures the ratio of the desorption rate to the shear rate, and the bulk Peclet number $Pe = \dot{\gamma}R^2/D_p$ measures the ratio of convective to diffusive transport in the bulk. The adsorption number $k = r_{a,p}\tilde{c}_{p,e}/r_{d,p}$, where $\tilde{c}_{p,e}$ is the initial equilibrium bulk concentration in the soluble phase, is fixed by requiring $J_{ps} = 0$ at $t = 0$, yielding $k = \omega/(1-\omega) = 1$. The adsorption depth $h = \tilde{s}_0/(R\,\tilde{c}_{p,e})$ measures the ratio of interfacial to bulk surfactant capacity; it is determined by the Langmuir equilibrium as $h = r_{a,p}\tilde{s}_\infty(1-\omega)/(r_{d,p}R) = 0.5$, consistent with \cite{Teigen2011}. Results are presented at $t = 0$, $4$, and $8$. The three solubility scenarios examined here correspond to qualitatively different physical settings, as discussed in Section~\ref{sec:model_selection}: the insoluble case represents a surfactant with negligible bulk solubility ($Bi = 0$); the single-phase-soluble case is representative of a gas-liquid system or a liquid-liquid system with a strongly hydrophilic or lipophilic surfactant, where one phase does not participate in bulk transport; and the partially soluble case represents a liquid-liquid system with a surfactant of intermediate HLB, soluble in both phases.

\subsubsection{Insoluble surfactant}
\label{sec:flow_insoluble}
\begin{figure}[H]
\captionsetup{labelfont=bf,font=footnotesize}
    \centering
    \includegraphics[width=0.9\textwidth]{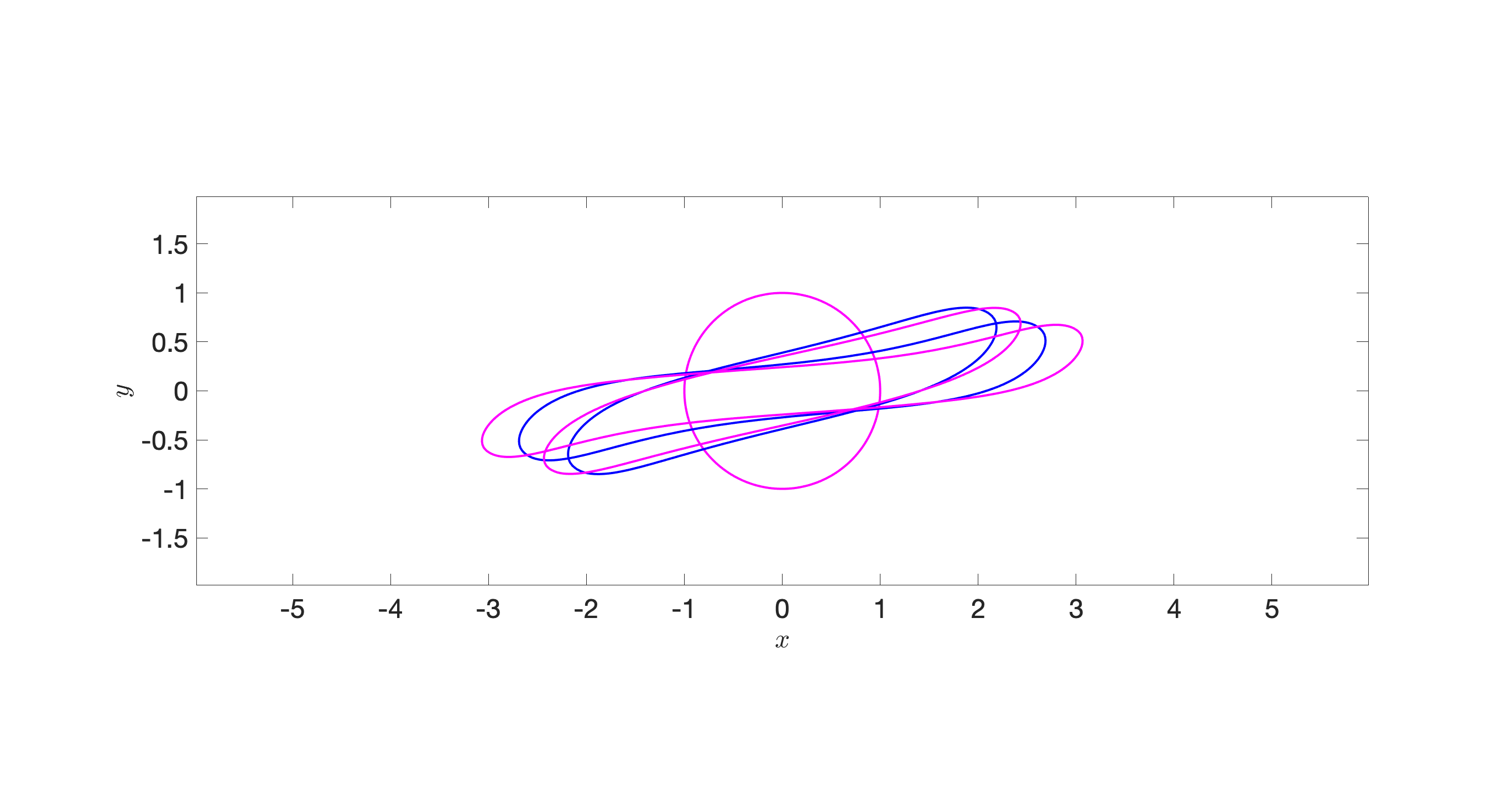}
    \caption{Drop shapes at $t = 0$, $4$, and $8$ for a clean drop (blue) and an insoluble surfactant-covered drop (magenta), following the setup of \cite{Teigen2011}.}
    \label{fig:flow_insoluble}
\end{figure}

Figure~\ref{fig:flow_insoluble} compares a clean drop with an insoluble surfactant-covered drop ($D_p = r_{a,p} = r_{d,p} = 0$) at $t = 0$, $4$, and $8$, reproducing the reference result of \cite{Teigen2011}. The surfactant-laden drop undergoes larger elongation and rotates faster than the clean drop, owing to the lower mean surface tension. As the drop deforms, the shear-induced surface circulation sweeps surfactant toward the tips of the elongating drop, where it accumulates and locally reduces $\sigma$, leaving the equatorial region with relatively higher $\sigma$. The resulting Marangoni stresses, directed from the tips (lower $\sigma$) toward the equatorial region (higher $\sigma$), oppose the surface circulation and partially retard the elongation. Our results are in good agreement with those reported in Fig.~7 of \cite{Teigen2011}.

\subsubsection{Surfactant soluble in the outer phase}
\label{sec:flow_soluble1}
\begin{figure}[H]
\captionsetup{labelfont=bf,font=footnotesize}
    \centering
    \begin{subfigure}{0.48\textwidth}
        \includegraphics[width=\linewidth]{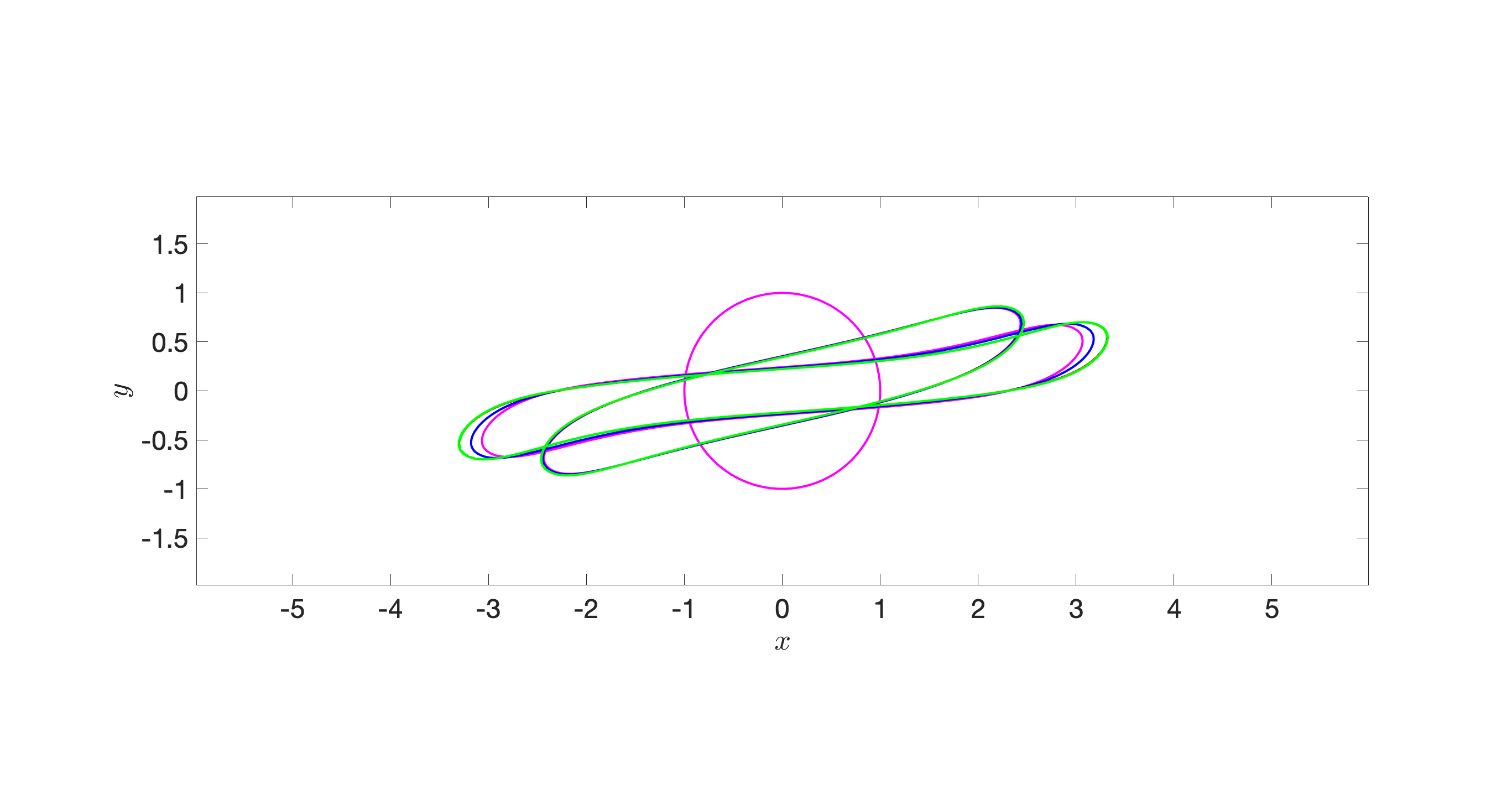}
        \caption{}
    \end{subfigure}
    \hfill
    \begin{subfigure}{0.48\textwidth}
        \includegraphics[width=\linewidth]{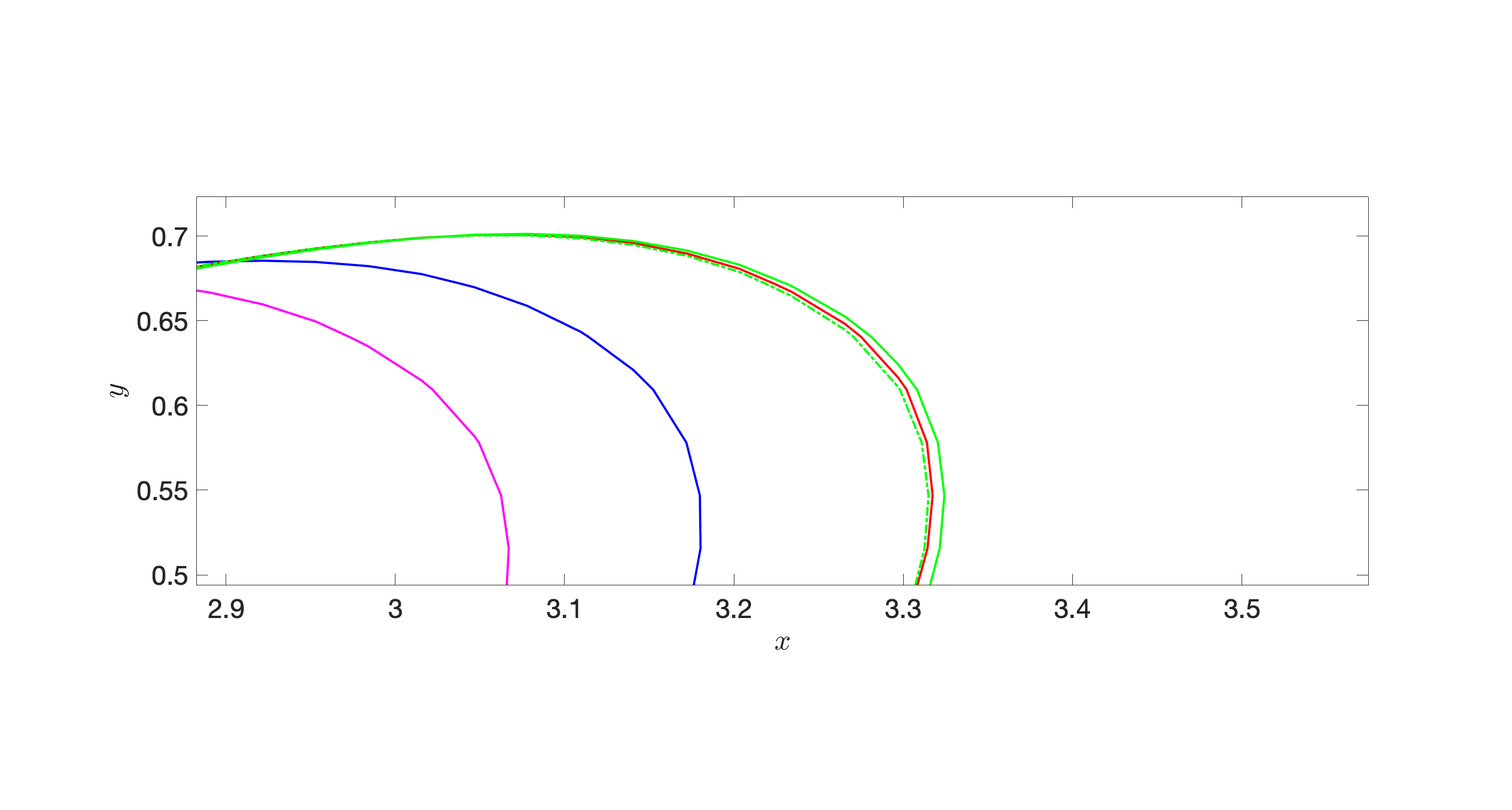}
        \caption{}
    \end{subfigure}
    \caption{Drop shapes at $t = 0$, $4$, and $8$ for a surfactant soluble in the outer phase (Section~\ref{sec:flow_soluble1}), varying the Biot number at fixed $Pe = 1$: insoluble (magenta), $Bi = 0.1$ (blue), $Bi = 1$ (red), $Bi = 10$ (green). Solid lines: three-scalar model; dashed lines: confined one-scalar model. (b) Close-up of the drop tip at $t = 8$.}
    \label{fig:soluble_Bi}
\end{figure}

\begin{figure}[H]
\captionsetup{labelfont=bf,font=footnotesize}
    \centering
    \begin{subfigure}{0.48\textwidth}
        \includegraphics[width=\linewidth]{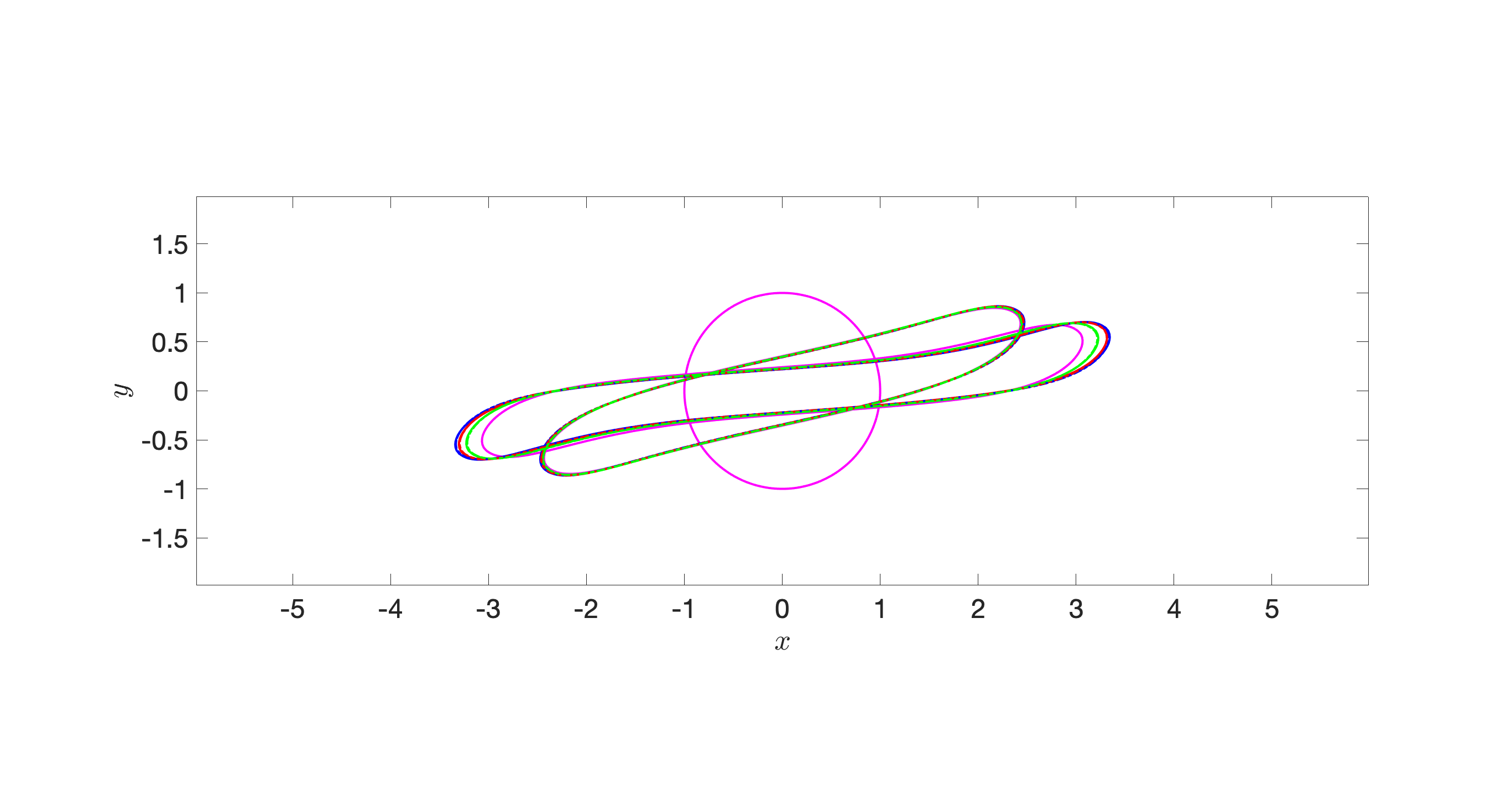}
        \caption{}
    \end{subfigure}
    \hfill
    \begin{subfigure}{0.48\textwidth}
        \includegraphics[width=\linewidth]{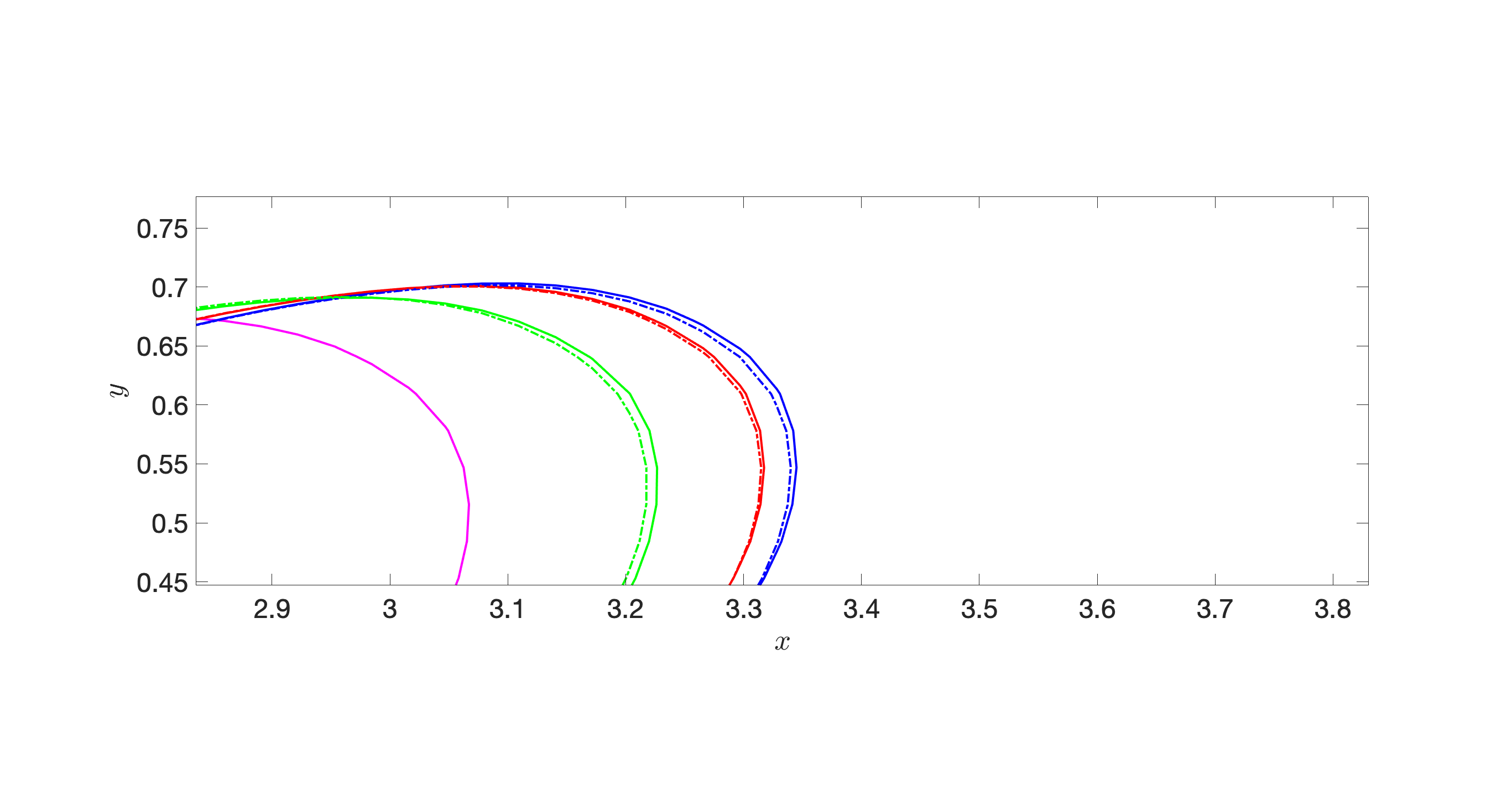}
        \caption{}
    \end{subfigure}
    \caption{Drop shapes at $t = 0$, $4$, and $8$ for a surfactant soluble in the outer phase (Section~\ref{sec:flow_soluble1}), varying the bulk Peclet number at fixed $Bi = 1$: insoluble (magenta), $Pe = 0.1$ (blue), $Pe = 1$ (red), $Pe = 10$ (green). Solid lines: three-scalar model; dashed lines: confined one-scalar model. (b) Close-up of the drop tip at $t = 8$.}
    \label{fig:soluble_Pe}
\end{figure}

We next consider a surfactant soluble only in the outer phase (phase~2), with $D_1 = r_{a,1} = r_{d,1} = 0$, following Section~5.2 of \cite{Teigen2011}. Figures~\ref{fig:soluble_Bi} and~\ref{fig:soluble_Pe} show the effect of varying the Biot number (at fixed $Pe = 1$) and the bulk Peclet number (at fixed $Bi = 1$), respectively, with solid lines for the three-scalar model and dashed lines for the confined one-scalar model.

Because the confined one-scalar model assumes instantaneous equilibrium between the interface and the bulk, the Biot number, which controls the rate of adsorption and desorption, drops out of its governing equations entirely. As a result, all one-scalar predictions in Figure~\ref{fig:soluble_Bi} collapse onto a single curve regardless of $Bi$; this curve represents the limit of infinitely fast exchange and is therefore closest to the three-scalar result at the largest $Bi$, where exchange is fastest. The bulk Peclet number, by contrast, controls bulk diffusion, which is retained in the one-scalar model; the one-scalar predictions in Figure~\ref{fig:soluble_Pe} therefore do depend on $Pe$, and the one-scalar and three-scalar models are in good agreement at each $Pe$ value, confirming that the equilibrium assumption is well-suited to this regime.

The Biot number governs the rate of exchange between the interface and the outer bulk relative to the shear rate. For small $Bi$, exchange is slow, the surface concentration remains non-uniform, and strong Marangoni stresses retard elongation. As $Bi$ increases, faster adsorption and desorption drive $\tilde{s}$ toward a more spatially uniform distribution, Marangoni stresses weaken, and the drop deformation increases, progressively approaching the equilibrium prediction. The bulk Peclet number has the opposite effect: decreasing $Pe$ corresponds to stronger bulk diffusion, which more rapidly homogenizes the bulk concentration adjacent to the interface, promoting surfactant exchange and driving $\tilde{s}$ toward uniformity. At large $Pe$, bulk transport is slow and the surface concentration gradients persist, sustaining significant Marangoni stresses. These trends are in good agreement with the results reported in Figs.~9--12 of \cite{Teigen2011}.

\subsubsection{Surfactant partially soluble in both phases}
\label{sec:flow_partial}
\begin{figure}[H]
\captionsetup{labelfont=bf,font=footnotesize}
    \centering
    \begin{subfigure}{0.48\textwidth}
        \includegraphics[width=\linewidth]{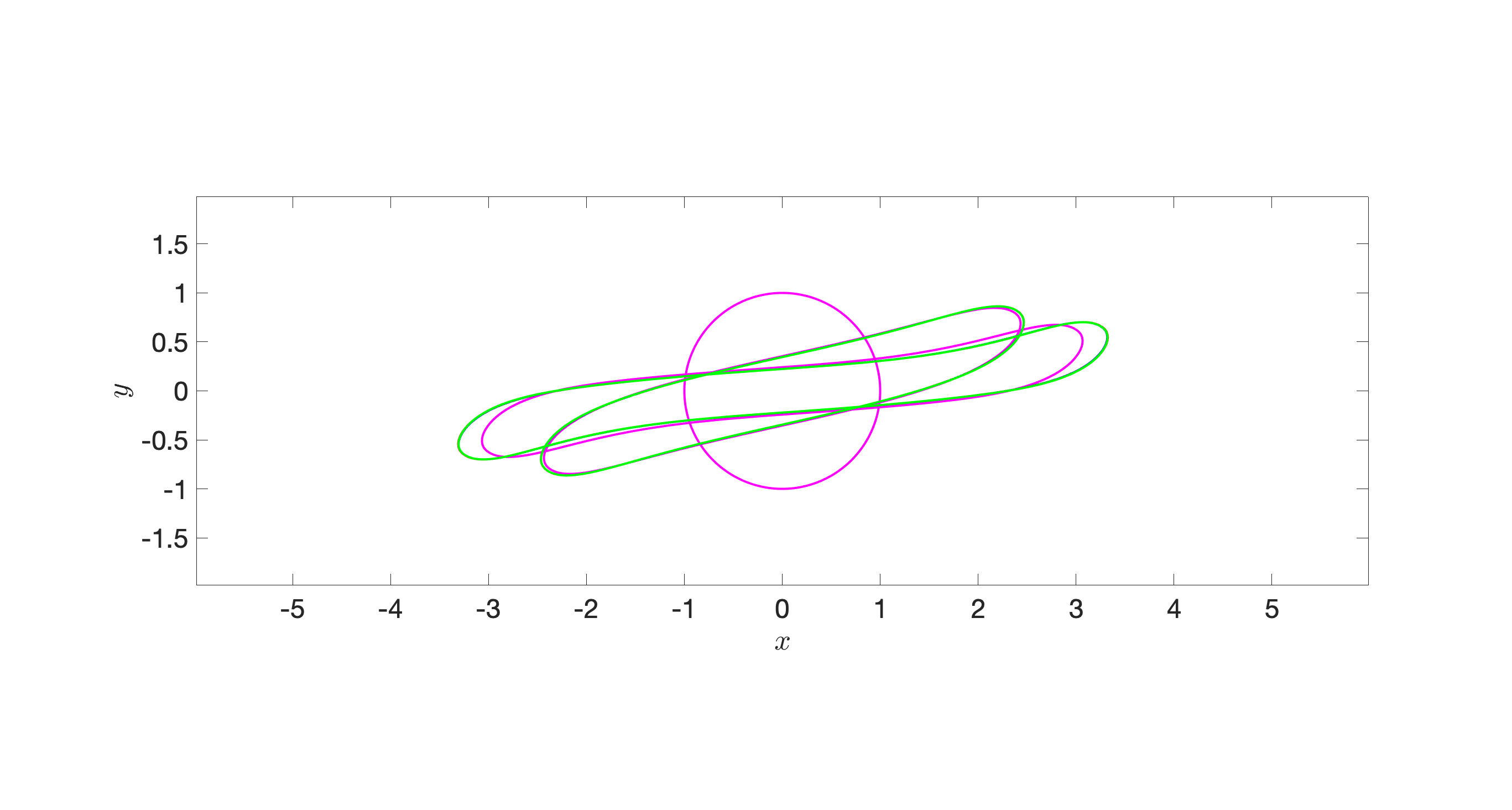}
        \caption{}
    \end{subfigure}
    \hfill
    \begin{subfigure}{0.48\textwidth}
        \includegraphics[width=\linewidth]{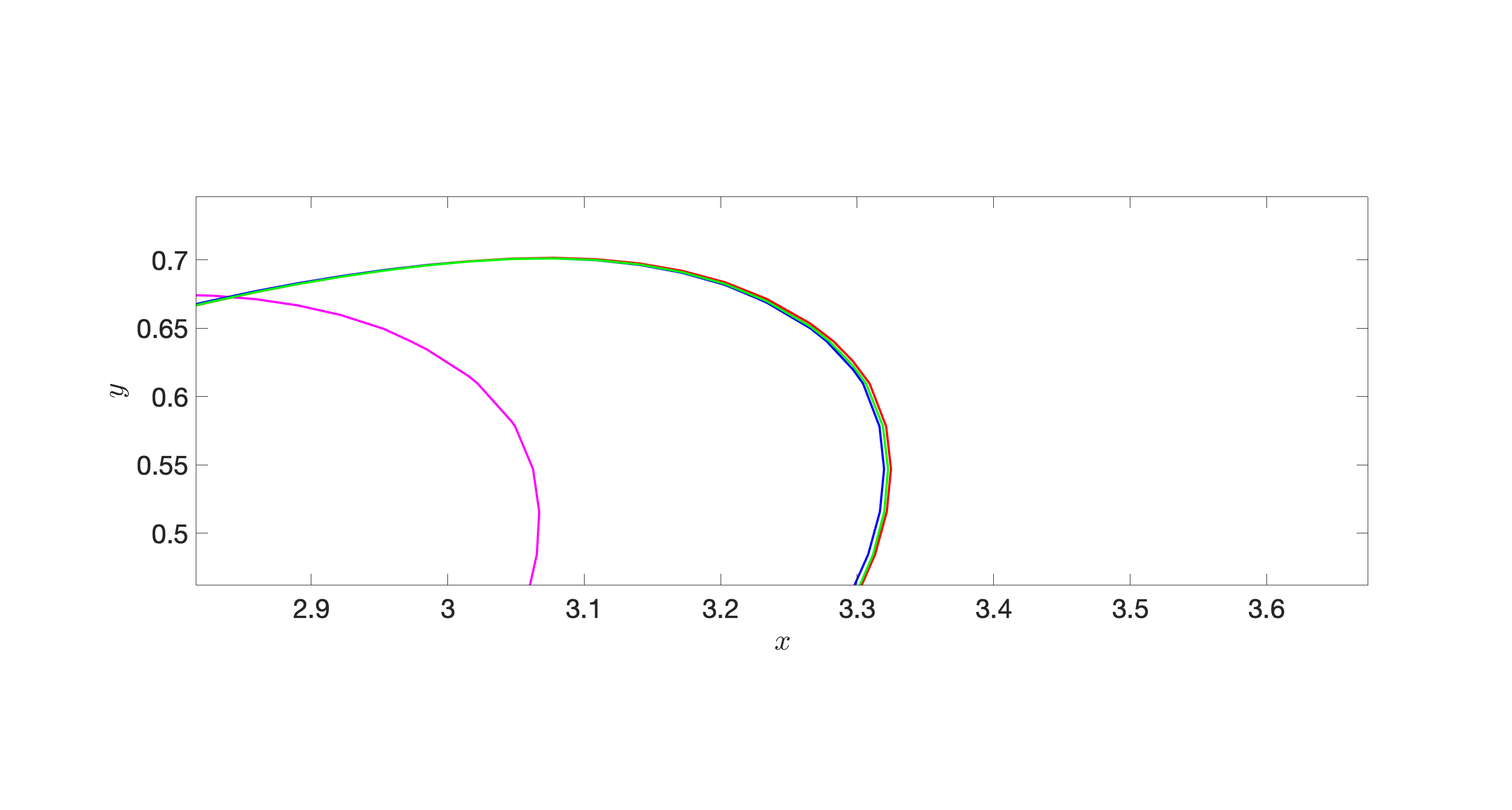}
        \caption{}
    \end{subfigure}
    \caption{Drop shapes at $t = 0$, $4$, and $8$ for a partially soluble surfactant (Section~\ref{sec:flow_partial}), varying the Biot number at fixed $Pe = 1$: insoluble (magenta), $Bi = 0.1$ (blue), $Bi = 1$ (red), $Bi = 10$ (green). (b) Close-up of the drop tip at $t = 8$.}
    \label{fig:partial_Bi}
\end{figure}

\begin{figure}[H]
\captionsetup{labelfont=bf,font=footnotesize}
    \centering
    \begin{subfigure}{0.48\textwidth}
        \includegraphics[width=\linewidth]{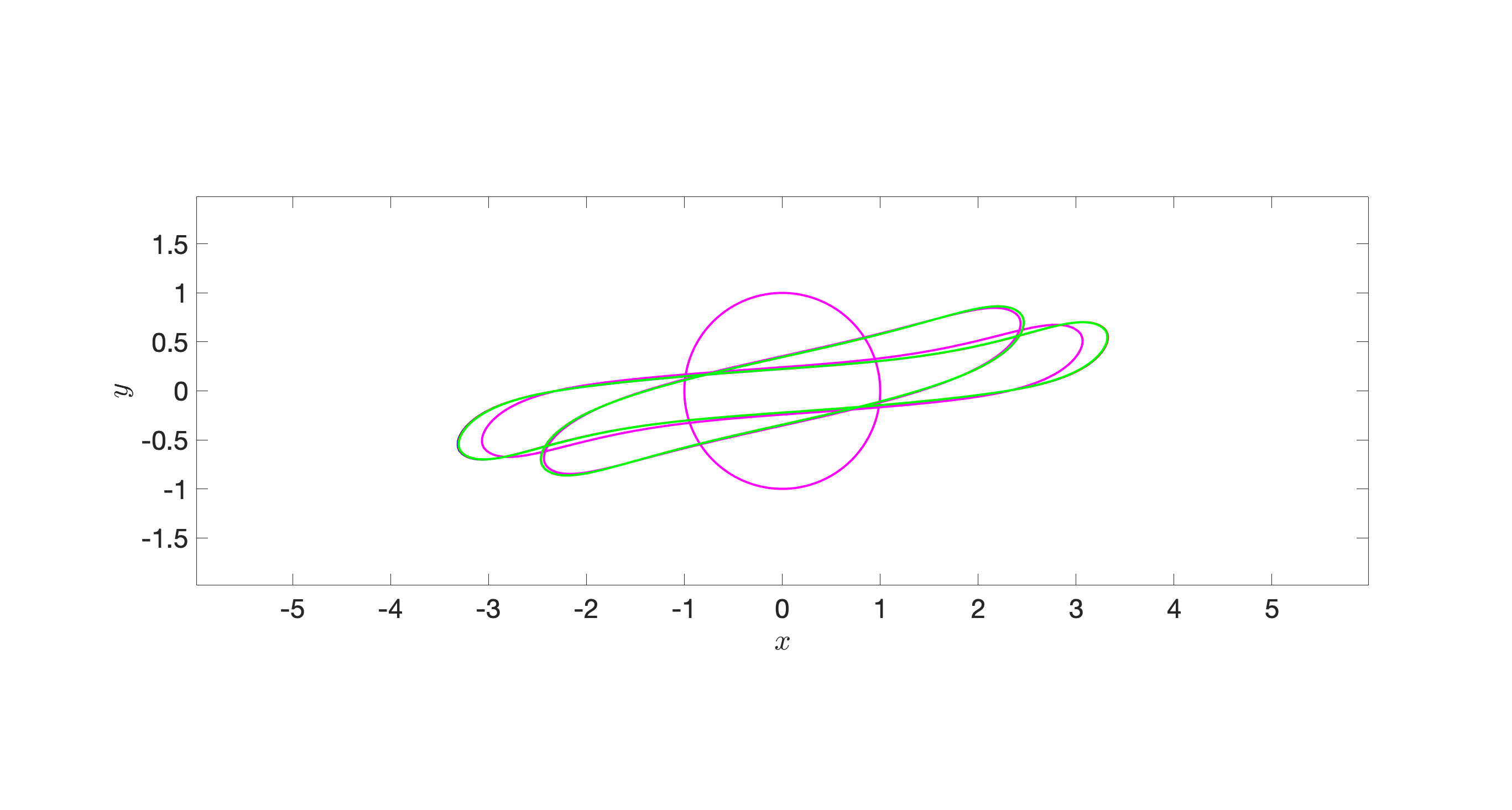}
        \caption{}
    \end{subfigure}
    \hfill
    \begin{subfigure}{0.48\textwidth}
        \includegraphics[width=\linewidth]{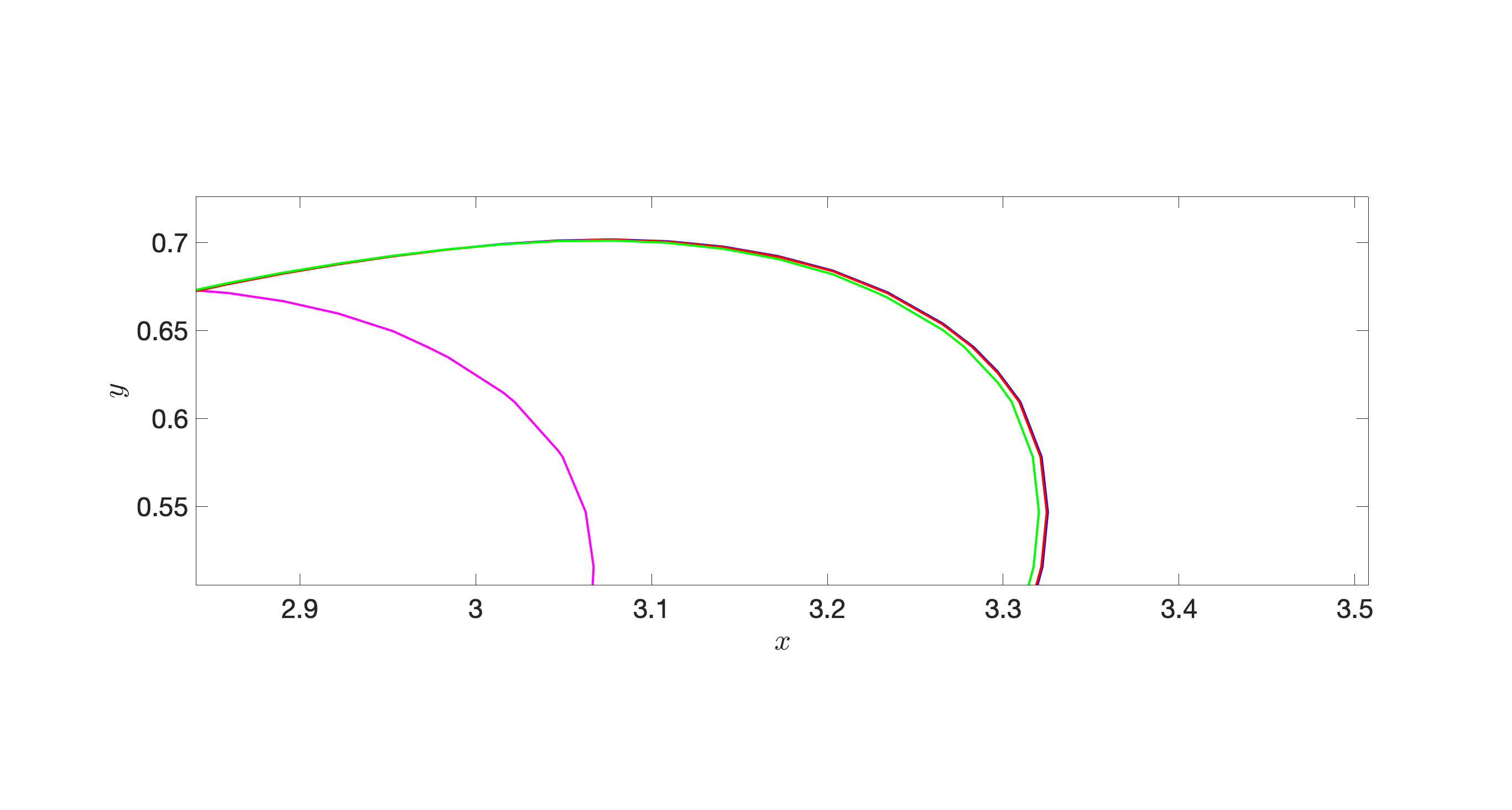}
        \caption{}
    \end{subfigure}
    \caption{Drop shapes at $t = 0$, $4$, and $8$ for a partially soluble surfactant (Section~\ref{sec:flow_partial}), varying the bulk Peclet number at fixed $Bi = 1$: insoluble (magenta), $Pe = 0.1$ (blue), $Pe = 1$ (red), $Pe = 10$ (green). (b) Close-up of the drop tip at $t = 8$.}
    \label{fig:partial_Pe}
\end{figure}

We now consider the partially soluble case, in which the surfactant exchanges with both the outer phase~2 and the drop interior (phase~1). This scenario has no counterpart in \cite{Teigen2011} and represents the novel contribution of this section. The surfactant properties in both phases are taken to be identical: $D_1 = D_2$, $r_{a,1} = r_{a,2}$, and $r_{d,1} = r_{d,2}$, so that the partition coefficient is $K_\text{eq} = 1$. No one-scalar model results are shown: unlike Section~\ref{sec:flow_soluble1}, there is no reference solution against which to assess the accuracy of the equilibrium reduction, and the focus here is a physics comparison between the partially soluble and single-phase-soluble cases rather than model validation. Figures~\ref{fig:partial_Bi} and~\ref{fig:partial_Pe} show the effect of varying $Bi$ (at fixed $Pe = 1$) and $Pe$ (at fixed $Bi = 1$), respectively, with the insoluble result (magenta) included as a reference.

The key difference from the single-phase-soluble case is that the drop interior now provides an additional exchange pathway for the surfactant. When a concentration gradient develops along the interface under shear, the surfactant can redistribute not only by desorbing into the outer bulk but also by transferring into and diffusing through the drop interior. This additional pathway further homogenizes $\tilde{s}$ along the interface, reducing Marangoni stresses compared to the single-phase-soluble case at the same $Bi$ and $Pe$, and leading to greater drop elongation. A striking consequence of this enhanced homogenization is that the results for different $Bi$ and $Pe$ values collapse nearly onto each other: with two exchange pathways active simultaneously, the surface concentration is driven toward uniformity even at moderate $Bi$ and $Pe$, leaving little sensitivity to the individual parameter values. The dominant effect is the contrast with the insoluble case, which exhibits significantly less elongation due to the absence of any bulk exchange.

\section{Summary}
\label{sec:conclusions}
We proposed a universal diffuse-interface modeling framework for surfactant transport in two-phase flows, applicable to all solubility scenarios and to any conservative phase field method. The foundation of the framework is a general three-scalar non-equilibrium model governing the surfactant concentrations in each bulk phase, $c_1$ and $c_2$, and at the interface, $s$. The model is locally and globally conservative, with the inter-phase transfer terms canceling pairwise so that the total surfactant amount $c_1 + c_2 + s$ is conserved in a closed domain. It is leakage-free, inheriting the property established in \cite{mirjalili2022computational,mirjalili2022consistent}: in the absence of physical adsorption and desorption, surfactant does not cross the interface unphysically even for large inter-phase concentration contrasts. The consistent inclusion of the phase field correction flux $\vec{R}$ ensures Galilean invariance, and the model is reduction-consistent: in the non-surfactant limit it recovers the interphase transfer model of \cite{mirjalili2022computational}, and in the limit of an artificial interface in a single-phase flow it recovers single-phase scalar transport. The framework is agnostic to the specific conservative phase field method employed, being applicable without modification to the Cahn-Hilliard equation, the conservative Allen-Cahn equation, and any other conservative formulation.

Assuming thermochemical equilibrium, we derived two one-scalar models from the three-scalar system: a full-equilibrium model for a surfactant in simultaneous equilibrium between both bulk phases and the interface, and a confined model for a surfactant in equilibrium between a single bulk phase and the interface. The three-scalar and one-scalar models are coupled to the Navier-Stokes equations through a surface tension force incorporating both the normal capillary term and the tangential Marangoni stress arising from non-uniform interfacial surfactant distributions.

We assessed the proposed models through a hierarchy of numerical tests. One-dimensional transport tests against analytical solutions confirmed that the three-scalar model accurately captures partially soluble, single-phase-soluble, and insoluble surfactant dynamics, and that it converges to the appropriate one-scalar equilibrium reduction as the system approaches thermochemical equilibrium. The confined one-scalar model was shown to be the appropriate reduction when one phase does not participate in transport, while the full-equilibrium model over-predicts surfactant concentrations in that regime. In the artificial-interface limit, the three-scalar model closely recovers single-phase scalar diffusion, with the small remaining difference vanishing both as $\epsilon\to0$ and at equilibrium. Two-dimensional advection-diffusion tests demonstrated convergence with mesh refinement for the soluble case, and agreement with an analytical solution for the insoluble case. Fully-coupled drop-in-shear flow simulations are in good agreement with the insoluble and single-phase-soluble results of \cite{Teigen2011}; the partially soluble case, for which no reference exists in the literature, is presented as a novel demonstration of the framework. For a surfactant soluble in the outer phase, we showed that the Biot number drops out of the confined one-scalar model — all one-scalar predictions collapse onto a single curve — while the three-scalar model captures the full Bi-dependence; the equilibrium reduction is approached only at large $Bi$. The partially soluble case, in which the surfactant exchanges with both the outer bulk and the drop interior, represents the novel contribution of this section: the additional exchange pathway further homogenizes the interfacial surfactant distribution, reduces Marangoni stresses, and leads to greater drop elongation compared to the single-phase-soluble case at equivalent parameter values.

\section*{Acknowledgments}
We acknowledge start-up funding from KTH Royal Institute of Technology and the Swedish e-science Research Center.

\bibliography{mybibfile}

\end{document}